\documentclass[prc,twocolumn,showpacs,amsmath,amssymb,floatfix,superscriptaddress,nofootinbib]{revtex4-1}
\usepackage{graphicx,graphics} 
\usepackage{amsmath}
\usepackage{amssymb}
\usepackage{epsfig}
\usepackage{epstopdf}
\usepackage{amstext}
\usepackage{amsthm}
\usepackage{amsfonts}
\usepackage{latexsym}
\usepackage{array}
\usepackage{xfrac}
\usepackage{color}
\usepackage{subfigure}
\usepackage{multirow}
\usepackage{fontenc}
\usepackage{bm}
\usepackage{adjustbox}
\usepackage{dcolumn}
\usepackage{ulem}
\begin{document}

\title{Microscopic-Macroscopic Approach for Ground-State Energies Based on the Gogny 
Force with the Wigner-Kirkwood Averaging Scheme}

\author{A. Bhagwat}
\affiliation{DM-DAE Centre for Excellence in Basic Sciences, Mumbai 400 098, India}
\author{M. Centelles}
\affiliation{Departament de F\'isica Qu\`antica i Astrof\'isica and Institut de Ci\`encies del Cosmos, 
Facultat de F\'isica, Universitat de Barcelona,  Mart{\'i} i Franqu{\`e}s 1, E-08028 Barcelona, Spain}
\author{X. Vi\~nas}
\affiliation{Departament de F\'isica Qu\`antica i Astrof\'isica and Institut de Ci\`encies del Cosmos, 
Facultat de F\'isica, Universitat de Barcelona, Mart{\'i} i Franqu{\`e}s 1, E-08028 Barcelona, Spain}
\author{P. Schuck}
\affiliation{Universit\'e Paris-Saclay, CNRS, IJCLab, IN2P3-CNRS, 91405 Orsay, France\\
Universit\'e Grenoble Alpes, CNRS, LPMMC, 38000 Grenoble, France}

\date{\today}

\begin{abstract}

In the previous paper I \cite{bhagwat20} we have shown that self-consistent Extended Thomas-Fermi (ETF) 
potentials and densities associated with a given finite-range interaction can be parametrized 
by generalized Fermi distributions. As a next step, a 
comprehensive calculation of ground-state properties of a large number of spherical and deformed even-even 
nuclei is carried out in the present work using the Gogny D1S force within the ETF scheme. The parametrized 
ETF potentials and densities of paper I are used to calculate the smooth part of the energy and the shell 
corrections within the Wigner-Kirkwood semiclassical averaging scheme. It is shown that the shell 
corrections thus obtained, along with a simple liquid drop prescription, yield a good description of 
ground-state masses and potential energy surfaces for nuclei spanning the entire periodic table.
\end{abstract}
\maketitle
\section{Introduction}
The development of radioactive beam facilities, such as Spiral, REX-Isolde, FAIR and the
future FRIB, has allowed to produce and determine the masses of many nuclei far away of the
stability line \cite{audi12}. Therefore, the study of the nuclear masses continues to be an
important and active field in nuclear physics. On the theoretical side there are basically
two different approaches to compute nuclear masses. One of them starts from 
effective interactions, such as the Skyrme \cite{vautherin72,guo91,chabanat98,stone07}, Gogny
\cite{decharge80,berger91,chappert08,goriely09}, Simple Effective Interaction (SEI) \cite{behera16} 
or M3Y \cite{nakada03} forces, energy density
functionals like BCPM \cite{baldo13,baldo17} or relativistic mean field models (RMF) \cite{RMF},
and calculates the masses through the Hartree-Fock-Bogoliubov (HFB) method \cite{ring-schuck}
with eventual additional corrections beyond mean field. As examples of HFB calculations of nuclear 
masses along the whole periodic table we shall mention the ones obtained by 
the Brussels-Montreal group (see \cite{goriely09a,goriely13} and references therein). 
They build up a sophisticated energy density functional using a generalizad
Skyrme interaction, which contains a density-dependent momentum term, a microscopic pairing contribution and
a macroscopic Wigner term. This functional depends on thirty parameters, which are fitted to reproduce, among another properties,
 2353 nuclear masses as well as the behaviour of microscopic equations of state in neutron matter at high density.
As a result, the so fitted BSk22$\to$Bsk26 forces predict  mass {\it rms} deviations between 629 and 544 keV \cite{goriely13}, which are 
among the best estimates of nuclear masses. Other parametrizations of the Skyrme force applied to 
large scale calculations of nuclear masses are the ones developed by the UNEDF collaboration. 
In Ref.~\cite{kortelainen14} they report three different fits of the Skyrme interaction to nuclear masses
that for 555 even-even nuclei predict {\it rms} deviations of 1.428 MeV (UNEDF0), 1.912 MeV (UNEDF1) and 1.950 MeV (UNEDF2),
which are in harmony with other estimates of masses for the same set of nuclei using different mean field models
\cite{baldo13}. Finally, let us mention that there exists another large-scale compilation of nuclear masses computed
through HFB calculations with the Gogny D1S force \cite{hilaire08}. 

Another method to obtain nuclear masses is the so-called Microscopic-Macroscopic (Mic-Mac)
model \cite{moller95,moller97,pomorski03,myers69,bhagwat10,bhagwat12}. This model is based on
the Strutinsky Energy theorem. According to this theorem, the nuclear ground-state energy
can be decomposed in two parts. One of them, which varies smoothly with  mass and atomic
numbers, is usually evaluated through liquid drop models of different degrees of sophistication.
The other part is an oscillating contribution, which is directly related to the quantal shell
effects. This part is the sum of the shell correction energy and the pairing correlation energy.
In the Mic-Mac models the oscillatory contribution is usually evaluated by means of an external
potential. The shell correction for each kind of particles is obtained as the total quantal energy 
(the sum of neutron and proton eigenvalues) minus the corresponding neutron or proton averaged energy. 
The fact that in Mic-Mac models the macroscopic part and the microscopic one are not connected  may be
 seen as a drawback if one has in mind an {\it ab initio} calculation with no ambiguities concerning the inputs 
and the results. However, in nuclear physics we are yet to reach a state where unambiguous {\it ab initio}
calculations can be done for any system across the periodic table. We have recourse to phenomenological 
density functionals, which are not unique and give the macroscopic properties in a somewhat 
indirect way. Hence, an independent but rather direct fit via a Liquid Drop Model (LDM) approach of the 
macroscopic properties can eventually have some practical
advantages. Energy density functionals derived from effective forces probably never can catch all the
correlations necessary for a fine tuning of the energies. The LDM has the advantage of 
fitting directly relevant quantities such as binding energies, surface and curvature energies, etc. 
For example, the theoretically difficult zero-point energies of HFB calculations are also directly fitted
into the LDM ones. On the contrary, in shell corrections, since they are obtained from 
differences of two energies, errors may cancel out. It is worth noting that succesful functionals such as those derived in
Refs.~\cite{goriely09a,goriely13} include a Wigner term, which has a clear macroscopic LDM origin.
Certainly the fine structure of the microscopic energy depends sensitively on details like spin-orbit, isospin dependence
of the mean field potential, effective mass, pairing correlations, etc. However, these contributions can be investigated
separately without the heavy machinery of fitting all the parameters from self-consistent calculations. It should also
be mentioned that there is still room for additional improvements of the Mic-Mac model, as for example in what concerns
the most important contributions around magic and doubly magic nuclei. In this sense we will pursue in this work our 
studies of the Mic-Mac model, which in any case is also among the more successful methods for predictions of nuclear masses.  \\
In many Mic-Mac calculations the average shell energy is obtained using the so-called Strutinsky method 
 \cite{strutinsky67,bunatian72}, which is a well-defined mathematical procedure for dealing
with the smoothing of shell effects in finite nuclei. This technique,
however, runs in practical difficulties for finite potentials because its calculation requires
the knowledge of the discrete single-particle spectrum at least in three major shells above the
Fermi level. For realistic nuclear single-particle potentials to perform the Strutinsky average
implies to take into account the continuum, which in many cases is discretized by diagonalizing
the single-particle Hamiltonian in a basis of an optimal size. This being a very delicate process 
and not everyone can handle this easily (see for example in this respect Ref.~\cite{kleban02})

A possible way to deal with the shell corrections bypassing these difficulties of the genuine Strutinsky 
smoothing is the so-called Strutinsky integral method \cite{chu77}. In this approach
one first minimizes the semiclassical energy density corresponding to a given effective interaction at Thomas-Fermi (TF)
or Extended Thomas-Fermi (ETF) level. Next, one computes the shell correction as the difference between the quantal energy
(sum of the eigenvalues of the lowest occupied single-particle levels) and the corresponding semiclassical counterpart
within the self-consistent TF or ETF mean field, considered as an external potential. In this way the microscopic energy
is added perturbatively to the macroscopic LDM energy provided by the semiclassical energy. A first estimate of nuclear 
masses using this method was perfomed in the eighties \cite{dutta86,tondeur87}. Later on a more complete mass table, 
called  ETFSI-I, was reported \cite{pearson91,aboussir92,aboussir95}. Later on a set
of mass formulae, computed also with the same method together with the MSk1$\to$MSk6 Skyrme forces,
was also obtained finding a mass {\it rms} in the range between 0.709 and 0.848 MeV \cite{tondeur00}. In addition, 
a study of the fission barriers of neutron-rich and superheavy nuclei was performed also using the ETF-plus-Strutinsky 
integral method \cite{mamdouh01}. This ETF-plus-Strutinsky integral method has been widely used in the context of 
neutron stars calculations to compute the EOS of the inner crust using Skyrme forces (see \cite{pearson18} and 
references therein) and Gogny interactions \cite{mondal20}.

To avoid this problem, we proposed some years ago an alternative technique. In Ref. \cite{bhagwat10}
it was shown that in order to evaluate the average energy of a set of $N$ neutrons and $Z$
protons in an external single-particle potential, the Strutinsky average could be replaced
by the corresponding semiclassical energy obtained by means of the Wigner-Kirkwood
(WK) $\hbar$-expansion of the one-body partition function
\cite{wigner32,kirkwood33,jennings75,ring-schuck,brack85,brack97,centelles06,centelles07}
associated to the external potential.
There are several reasons supporting this choice of using the WK approach instead of the
Strutinsky average to compute the shell corrections. First, the Strutinsky level
density is known to be an approximation to the Wigner-Kirkwood level density in the least 
square sense \cite{azizi06}.
Second, the WK level density avoids the problems related with the treatment of
the continuum, as far as the upper limit of the energy needed for its calculation is the Fermi
level. Although the WK level density including $\hbar$-corrections, $g_{\rm WK}(\varepsilon)$, has a
$\varepsilon^{-1/2}$ divergence for potentials that vanish at large distances \cite{shlomo91},
the integrated moments of this quantity are well behaved \cite{centelles07}.

As it has been shown in previous literature \cite{bhagwat10,bhagwat12}, the use of the WK
expansion to compute the shell correction allows one to obtain ground-state masses along the whole periodic
table with a quality similar to that found using the well-established Mic-Mac models such as the
FRDM of M\"oller-Nix \cite{moller97} or the Lublin-Strasbourg Drop (LSD) \cite{pomorski03}
 for the same set of nuclei. In this
work we explore the interesting possibility of using self-consistent single-particle potentials
computed with effective nuclear interactions to calculate shell corrections
using the WK approximation. This choice would provide a link between the well-known mean-field
approximation using effective forces and the Mic-Mac models. To this end, instead of the fully
quantal single-particle potential obtained from the Hartree-Fock-Bogoliubov (HFB) scheme, we
use, in the spirit of the so-called expectation value method \cite{brack85,bohigas76}, 
the semiclassical single-particle potentials developed in Paper I \cite{bhagwat20}. These potentials have 
been calculated self-consistently in the Extended Thomas-Fermi (ETF) approximation (see Paper I for 
further details). The ETF approach is based on the
WK expansion of the distribution function and has some advantages. First, due to the fact that
the  ETF method is deeply rooted in the classical periodic orbit theory \cite{brack97}, it gives
a very intuitive picture of the physical process. Second, the ETF approach provides energy density
functionals expanded order by order in $\hbar$. The ETF approach has been widely used together
with Skyrme forces for describing binding energies of finite nuclei at zero and finite temperature 
\cite{brack85} as well as in the RMF framework \cite{centelles90,centelles93}.
The ETF approach has also been extended to the case of non-local single-particle 
Hamiltonians \cite{soubbotin00} and, therefore, can be applied to the case of effective finite range 
forces, like the Gogny interaction \cite{soubbotin00,gridnev98,soubbotin03} as we will do in this work.

We begin with a very brief overview of the essentials of the Mic-Mac approach 
using the WK averaging scheme. The results will be presented and 
discussed in the third and fourth sections. The summary and conclusions are contained in the last section.

\section{Formalism and Details of Calculations}
\subsection{The microscopic part of the model}
The essential ingredient to evaluate the microscopic part in the Mic-Mac models
is the external single-particle potential. Using this potential the quantal effects, namely the shell
corrections and the pairing correlations, are calculated. Usually this external mean-field
is chosen as a phenomenological potential that is able to reproduce as closely as possible the experimental
single-particle energy levels of some selected nuclei. Examples of these potentials are the one derived by 
Wyss that was used in our previous Mic-Mac WK calculations \cite{bhagwat10,bhagwat12} or the Yukawa-folded-interaction
used in the FRDM \cite{moller97} and LSD \cite{pomorski03} Mic-Mac calculations. Alternatively, in the
present work we propose to use as external potential the one obtained semiclasically employing the
D1S Gogny interaction. For practical purposes this Gogny-based potential has been fitted 
to generalized Woods-Saxon functions, as has been explained in detail in the previous Paper I
\cite{bhagwat20}. Our aim here is to see, on the one hand, whether, and in which way, this procedure 
can compete with the version where also the mean field potential is entirely phenomenological. On the other hand, 
we want to investigate what can be learnt from the comparison between our Mic-Mac and the HFB calculations 
obtained with the same interaction.

The single-particle Hamiltonian reads
\begin{eqnarray} 
\hat{H}~=~\frac{-\hbar^2}{2m} \nabla^2 ~+~V(\vec{r})~+~\hat{V}_{LS}(\vec{r}),
\label{eq1}
\end{eqnarray} 
where $V(\vec{r})$ is the one-body central potential and $\hat{V}_{LS}(\vec{r})$ the 
spin-orbit potential.
In order to remain as close as possible to the phenomenological mean-field potentials, we chose
the semiclassical Gogny single-particle potential derived from the energy density that
includes the effective mass contribution in the potential energy part (see Paper I for further details),
which is consistent with the single-particle Hamiltonian (\ref{eq1}).

As discussed in Paper I, the central and spin-orbit potentials entering in (\ref{eq1})
are computed at ETF level with the D1S Gogny interaction and then parametrized for each type of nucleon $i=n,p$ as 
\begin{eqnarray}
V^{(i)}(r) = V_{m}^{(i)}\left[f\left(r;R_{m}^{(i)},a_{m}^{(i)}\right)\right]^{\nu_{m}^{(i)}}
\label{eq2}
\end{eqnarray}
and 
\begin{eqnarray}
V^{(i)}_{so}\left(r\right) = U_{so}^{(i)}\frac{d}{dr}\left\{
\left[f\left(r;R_{so}^{(i)},a_{so}^{(i)}\right)\right]^{\nu_{so}^{(i)}}\right\},
\label{eq3}
\end{eqnarray}
respectively, with
\begin{eqnarray}
f(r)~=~\frac{1}{1 + \exp{\left[l(\vec{r})/a\right]}}~.
\label{eq4}
\end{eqnarray}
In Eqs.~(\ref{eq2}) and (\ref{eq3})  $V_{m}^{(i)}$ and  $U_{so}^{(i)}$ are the 
strengths of the central and spin-orbit potentials, respectively, $a$ is the
diffuseness and $l(\vec{r})/a$ the distance function, which is defined under
the requirement that the skin thickness remains constant through the 
nuclear surface. Thus the distance function reads \cite{bhagwat10}
\begin{eqnarray}
l(\vec{r})~=~\frac{r~-~R_s}{|\nabla{(r~-~R_s)}|_{r=R_s}}~,
\label{dist}
\end{eqnarray}
where $R_s$ is the position of the deformed surface, which is parametrized as 
\begin{eqnarray}
R_s~=~CR_0\left(1~+~\sum_{\lambda,\mu}\alpha_{\lambda,\mu}Y_{\lambda,\mu}\right)~.
\label{def}
\end{eqnarray}
In this equation $R_0$ is the half-density radius of the Woods-Saxon
function (\ref{eq2}), $C$ is a constant to ensure the volume conservation 
and the coefficients $\alpha_{\lambda,\mu}$ are related to the three degrees
of freedom considered in this work, namely $\beta_2$, $\beta_4$ and $\gamma$,
through the standard relations given in \cite{bhagwat10}. 
The numerical values of the parameters that define the central (Eq. (2)) and spin-orbit (Eq. (3))
potentials  are reported in Appendix~2 of Paper~I.

As we did in previous calculations \cite{bhagwat10,bhagwat12}, we compute the Coulomb 
potential, which  for protons contributes to the central part of the single-particle Hamiltonian 
(\ref{eq1}), by folding the proton density with the Coulomb interaction. In order to simplify 
the calculation, we take the proton density as parametrized in Paper I and use the same deformation 
parameters as for the nuclear potential for protons.

The shell correction is given by the difference between the quantal energy and its 
averaged counterpart. In the case of an external potential the quantal energy is given by the
sum of eigenvalues associated to the single-particle Hamiltonian (\ref{eq1}).
The average energy in our Mic-Mac model is given by the WK energy  associated to
the Wigner transform  of the quantal Hamiltonian (\ref{eq1}).
The pairing correlations are important for open-shell nuclei.
As we have done in our previous works \cite{bhagwat10,bhagwat12}, we include the pairing effects
for both neutrons and protons through the Lipkin-Nogami model of pairing on top of the quantal
Hamiltonian (\ref{eq1}).
The microscopic energy, which  in the Mic-Mac models is given by the sum of the shell corrections
and pairing correlations for each type of nucleons, reads
\begin{eqnarray}
\delta E  = E_{shell,n} + E_{pair,n} + E_{shell,p} + E_{pair,p}.
\label{eq5}
\end{eqnarray}

 To obtain the semiclassical WK energy the starting point 
is the quantal partition function,
\begin{eqnarray}
Z\left(\beta\right)~=~\mathrm{Tr}\left(\exp{(-\beta\hat{H})}\right), 
\end{eqnarray} 
where  $\hat{H}$ is the Hamiltonian of the system  (\ref{eq1}), 
which includes the central and spin-orbit terms.

The semiclassical WK expansion of the one-body partition function in powers of Planck's 
constant $\hbar$ was developed by Wigner \cite{wigner32} and 
Kirkwood \cite{kirkwood33}. It allows one to obtain systematic corrections to the Thomas-Fermi 
energy and particle number (see for example also Refs.~\cite{jennings75,ring-schuck,brack97,centelles07} for more
details). Here, we expand the semi-classical partition function up to the fourth order in $\hbar$. 
Symbolically, this can be expressed as:
\begin{eqnarray} 
Z_{\rm WK}^{(4)}(\beta)~=~Z^{(4)}_{\rm CN} (\beta)~+~Z^{(4)}_{\rm SO}(\beta) \,,
\label{ZWK}
\end{eqnarray} 
where
$Z^{(4)}_{\rm CN}(\beta)$ and $Z^{(4)}_{\rm SO}(\beta)$ are the WK partition functions 
for the central and spin-orbit terms~\cite{jennings75}, respectively. 
For each kind of nucleons, the level density, energy, and particle number can be obtained through suitable 
Laplace inversions of the partition function as follows:
\begin{eqnarray}
g_{\rm WK}(\epsilon)~=~{\cal{L}}^{-1}_{\epsilon} Z_{\rm WK}^{(4)}(\beta)~,
\end{eqnarray}
\begin{eqnarray}
N~=~{\cal{L}}^{-1}_{\lambda} \left( \frac {Z_{\rm WK}^{(4)}(\beta)} {\beta} \right) ,
\label{N_qm}
\end{eqnarray}
and
\begin{eqnarray}
E_{\rm WK}~=~\lambda N~-~{\cal{L}}^{-1}_{\lambda} \left( \frac {Z_{\rm WK}^{(4)}(\beta)} {\beta ^2} \right) ,
\label{E_qm}
\end{eqnarray}
where $\lambda$ is the chemical potential, fixed by demanding the right
particle number. Details of this procedure as well as 
the corresponding formulas for the various quantities can be found in Refs.~\cite{bhagwat10,bhagwat12}.

According to Ref.~\cite{jennings75}, we write the WK energy in the following way:
\begin{eqnarray} E_{\rm WK} &=& \lambda N - \left( 
E_{\hbar^{0}}^{\rm CN} + E_{\hbar^{2}}^{\rm CN} + E_{\hbar^{4}}^{\rm CN} \right)
           - \left( E_{\hbar^{2}}^{\rm SO} + E_{\hbar^{4}}^{\rm SO} \right),
\nonumber \\
\label{EJEN}
\end{eqnarray}
where $E_{\hbar^{2k}}^{\rm CN}$ and $E_{\hbar^{2k}}^{\rm SO}$
denote the contributions to the average energy of the order
$\hbar^{2k}$ arising from Laplace inversion of the central and spin-orbit
parts of the partition function (\ref{ZWK}), respectively. Explicit expressions
of each contribution to the WK energy in Eq.~(\ref{EJEN}) are reported in
Ref.~\cite{bhagwat12} and we summarize them in Appendix~1 for the sake of completeness.

\subsection{The macroscopic part of the model}

The macroscopic part of the energy is determined using the liquid drop model. Here,
we use a version inspired by the one of Dudek and Pomorski \cite{bhagwat10,bhagwat12,pomorski03}
\begin{widetext}
\begin{eqnarray}
E_{mac}&=&a_v\left[1+\frac{4k_v}{A^2}T_z\left(T_z+1\right)\right]A
      + a_s\left[1+\frac{4k_s}{A^2}T_z\left(T_z+1\right)\right]A^{2/3} \nonumber \\
&+& a_{cur}\left[1+\frac{4k_{cur}}{A^2}T_z\left(T_z+1\right)\right] A^{1/3}
       + \frac{3Z^2e^2}{5r_0A^{1/3}}+\frac{C_4Z^2}{A} + E_{W} ,
\label{Emac}
\end{eqnarray}
\end{widetext}
where $a_v$, $a_s$, $a_{cur}$, $k_v$, $k_s$, $k_{cur}$, $r_0$ and $C_4$ are free parameters,
$T_z = \left|N - Z\right|/2$ is the third component of isospin, $e$ is  electronic charge and $E_{W}$ is 
the Wigner energy,
given by:
\begin{eqnarray}
E_{W}=w_{1}\exp\left\{-w_{2}\left| \frac{N-Z}{A}\right |\right\} \Theta\left(20-Z\right)
\Theta\left(40-A\right)\nonumber \\
\label{Ewig}
\end{eqnarray}
with $w_1$, $w_2$ as free parameters. Most of the nuclei considered in the
investigation are deformed. The liquid drop quantities defined above, in particular,
surface, curvature and Coulomb energies, therefore become deformation dependent.
Details can be found in~\cite{bhagwat12}.
It is important, however, to point out that in our previous works \cite{bhagwat10,bhagwat12}
the curvature energy was dropped because we had found that it was very difficult to adjust 
the corresponding parameter reliably: the {\it rms} error in the parameter worked out to be of
the order of 100\%. In the present investigation, however, the curvature correction is found 
to make a significant contribution. The inclusion of the curvature correction (without any 
isospin component) 
 is crucial to ensure that the isotopes $^{182,184,186}$Pb work out to be spherical (see below).

\subsection{The fitting procedure}

As we have mentioned before, the total energy in our Mic-Mac model can be written as
\begin{eqnarray}
E(N,Z,\hat{\beta}) = E_{LD}(N,Z,\hat{\beta}) + \eta \,\delta E(N,Z,\hat{\beta}) ,
\label{eq15}
\end{eqnarray}
with $E_{LD}(N,Z,\hat{\beta})$ being the LDM part of the energy and $\delta E(N,Z,\hat{\beta})$
the microscopic part of the energy (shell correction plus pairing {\it \`a la} Lipkin-Nogami). $N$ and $Z$
represent the neutron and proton numbers, and the symbol $\hat{\beta}$ stands for three deformation
parameters, namely, $\beta_2$, $\beta_4$ and $\gamma$. In our previous works \cite{bhagwat10,bhagwat12}, where we
developed the WK Mic-Mac model based on the Wyss potential, the renormalization factor $\eta$ was 
chosen to be 0.85 (see \cite{bhagwat10,bhagwat12} for details). 

In this work, where we use the Gogny-based mean-field as external potential, we proceed in a similar way. Starting 
with the value $\eta$=0.85 and from the optimal deformation parameters for each nucleus obtained from the Wyss potential
\cite{bhagwat12}, we have performed a new minimization for each considered nucleus in order to find the optimal 
deformation parameters associated to the Gogny-based single-particle potential as well as the coefficients of the 
macroscopic part Eq.~(\ref{Emac}). However, the Gogny Mic-Mac model fitted in this way ran into troubles while the 
potential energy surfaces were being explored. It turned out that the $^{182,184,186}$Pb isotopes had strong prolate 
minima, which is not acceptable, given that these are all semi-magic nuclei, and $Z=82$ is a very robust proton shell 
closure. It was also found that this lack of sphericity of the previously mentioned Pb isotopes was not avoided by 
varying the $\eta$ parameter within a range of reasonable values. To solve this problem it is important to point out 
that the deformation properties, which impact on the mic part, are strongly linked with the surface properties 
of the macroscopic part of the model, which are determined by the surface and curvature terms in Eq.~(\ref{Emac}). 
Therefore 
the pathology found in the shape of $^{182,184,186}$Pb should be attributed to a deficiency of the macroscopic part 
because 
only the surface contribution was taken into account in the minimization procedure. Thus, we have performed a new 
minimization of the difference between the theoretical and experimental energies by adopting $\eta$=0.67. In this 
minimization we have additionally explicitly taken into account the curvature coefficient, along with its deformation dependence, 
in the macroscopic part and checked explicitly that the nuclei $^{182,184,186}$Pb were spherical in their ground state.

The liquid drop parameters in (\ref{Emac}) for the Gogny-based WK Mic-Mac model fitted to the 
experimental energies \cite{audi12} (without the electronic binding energy, which has 
been subtracted from the energies reported in \cite{audi12}) of 551 even-even spherical and deformed nuclei are 
reported in 
Table \ref{Table1} with the label ``exp". The complete list of energies of these 551 nuclei can be found in the 
supplemental material \cite{online}. The rms deviation of the energies from experiment is of 834 keV, as 
reported in the bottom row of Table \ref{Table1}. For the sake of ascertaining whether our Mic-Mac approach 
leads to consistent results, we have also performed another different fit of the macroscopic 
part of our model to reproduce the Gogny-D1S HFB energies of the same set of nuclei. 
The values of the corresponding liquid drop parameters are given in the same Table with the label HFB; 
the energy rms deviation from experiment in this case increases to 3.95 MeV.
It is worthwhile to mention here that the isospin curvature 
term in (\ref{Emac}) is not considered in these fits for the following reasons. On the one hand, the statistical 
error of this term is usually very large. On the other hand, this term usually weakens the isospin surface term by a 
large factor. In our fit of the macroscopic part we have also included the Wigner term, which is relevant for 
describing 
light nuclei and practically not needed for nuclei with mass numbers greater than $A$=40. Our Gogny-based 
WK Mic-Mac model has been fitted for atomic numbers above $Z$=20, which implies that only few nuclei are affected by 
the Wigner term. We have checked that if this term is not taken into account in the macroscopic part of the energy,
one obtains basically the same energy {\it rms} deviation. The reason for that is a correlation between the 
curvature and Wigner terms, which produce larger contribution of the curvature energy when the Wigner term is not
considered. In the third column of Table \ref{Table1} we report with the label ``HFB" the liquid drop parameters 
of the model fitted to reproduce the HFB energies computed with the D1S Gogny interaction. In this case the fit is 
performed without taking into account the Wigner term by the reasons pointed out before.

\begin{table}[htb]
	\begin{center}
		\caption{Liquid drop parameters in Eqs.(\ref{Emac}) and (\ref{Ewig}).}\vspace{5pt}
\label{Table1}
\begin{tabular}{ l|c|c } \hline
Parameter           &   `exp'      &   `HFB'  \\ \hline
   $a_v$ (MeV)      & -15.9030   &  -16.7564  \\ 
   $k_v$            &  -1.8549   &   -1.9764  \\ 
   $a_s$ (MeV)      &  20.2648   &   26.2775  \\ 
   $k_s$            &  -2.1017   &   -2.1297  \\ 
   $a_{cur}$ (MeV)  &  -3.7770   &  -12.6004  \\  
   $k_{cur}$        &   0.~~~~   &    0.~~~~  \\
   $r_0$ (fm)       &   1.1919   &    1.1376  \\ 
   $C_4$ (MeV)      &   1.3210   &    2.0892  \\
   $w_1$ (MeV)      &  -1.5279   &            \\ 
   $w_2$            &   7.8563   &            \\
{\it rms} dev.~(MeV)&   0.834    &    3.950   \\ \hline
\end{tabular}
\end{center}
\end{table}

\section{Results and Discussions}

\subsection{Mic-Mac versus HFB calculations with the D1S Gogny interaction}

\begin{figure*}[htb]
\centering \includegraphics[scale=0.4]{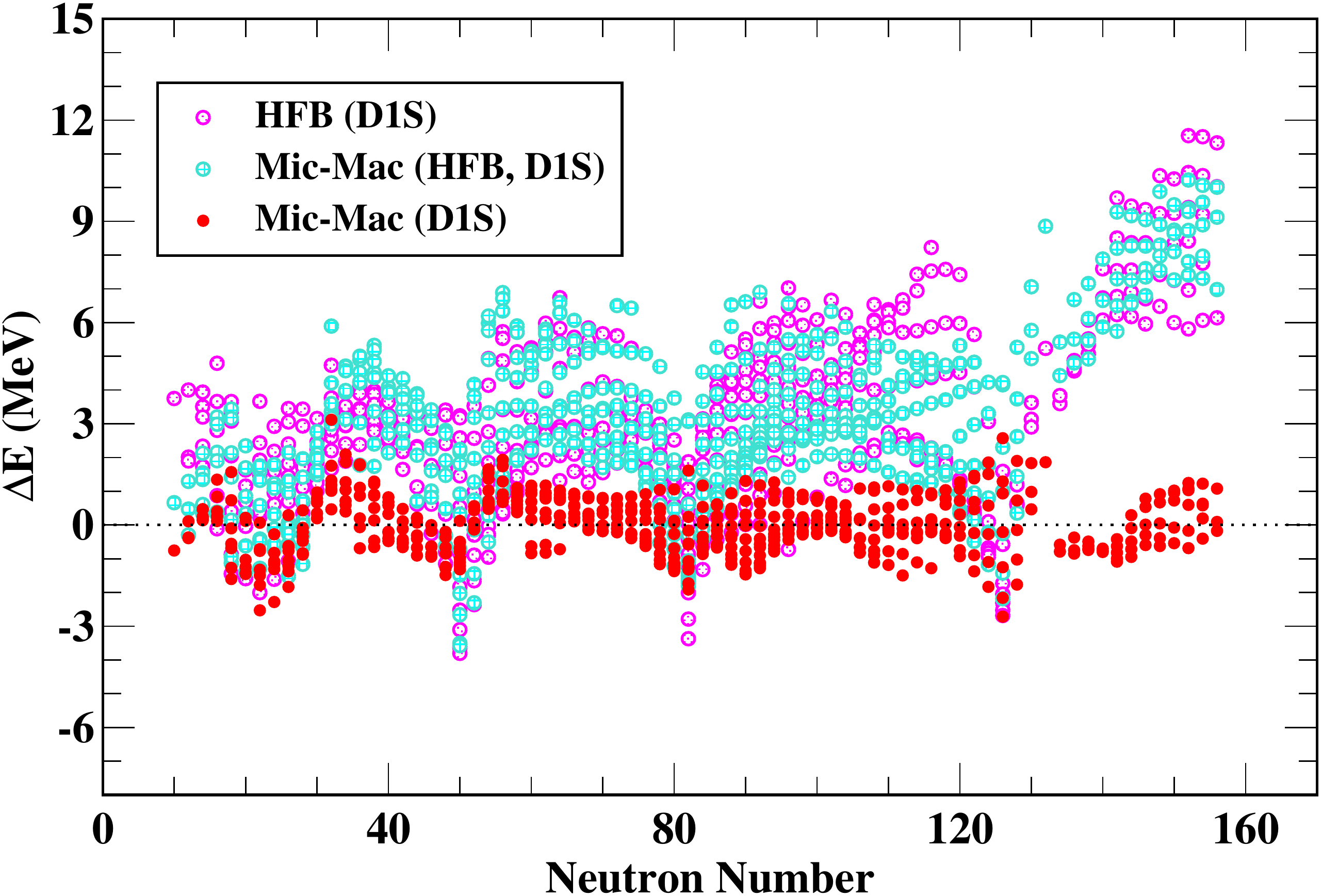}
\caption{Residues with respect to the experimental energies of 551 spherical and deformed even-even
nuclei predicted by the HFB calculation with the Gogny D1S force (magenta symbols)
\cite{luis} and by the Mic-Mac Gogny-based models fitted to 
experimental energies (red symbols) and to HFB energies (cyan symbols).}
\label{micmac_hfb}
\end{figure*}

\begin{figure*}[htb]
\centering \includegraphics[scale=0.4]{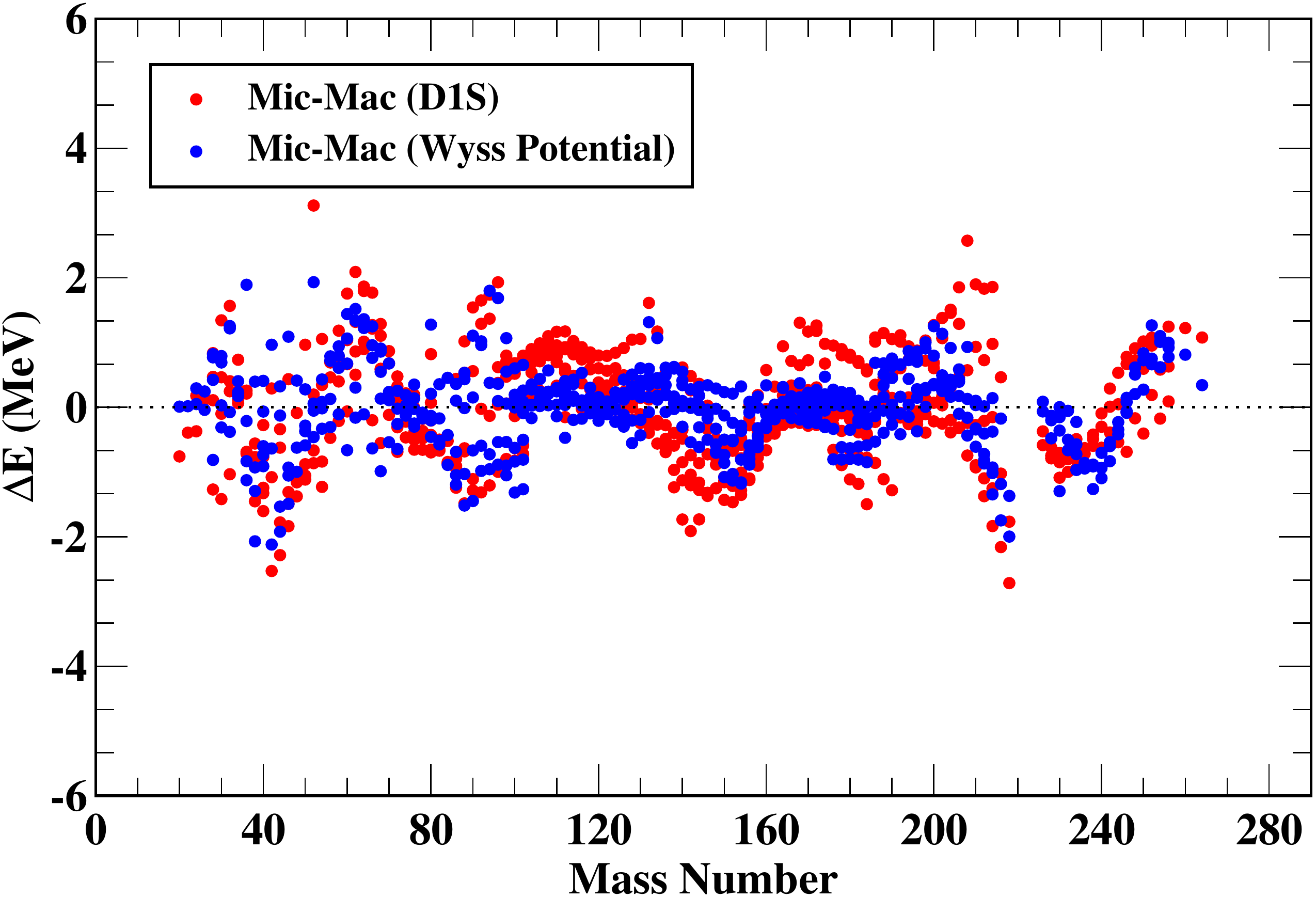}
\caption{Residues with respect to experiment for the energies of 551 spherical and deformed even-even nuclei
calculated with the WK Mic-Mac models based on the Gogny D1S force (red symbols) 
and on the phenomenological Wyss potential (blue symbols).}
\label{difbe}
\end{figure*}

In Figure \ref{micmac_hfb} we display by red symbols the residues (the differences
between the calculated and experimental energies) of 551 spherical and deformed even-even 
nuclei, the  masses of which are experimentally well determined, calculated with our 
Gogny-based WK Mic-Mac model. In the same figure we show 
with magenta symbols the residues of the HFB energies computed using the same D1S 
force and the same set of nuclei. As it can be seen, the pattern exhibited by the two calculations 
performed with the same D1S interaction is clearly different. 
On the one hand, the HFB energies calculated with the D1S interaction 
show the well-known energy drift for neutron-rich nuclei \cite{pillet17}, while in the 
Mic-Mac calculation, where the macroscopic part is fitted to experimental masses, 
this drift is completely washed out and the predictions are very similar 
to the ones obtained in our previous calculation \cite{bhagwat12} with the phenomenological 
Wyss potential, as it can be clearly appreciated in Figure \ref{difbe}. On the other hand, 
our Gogny-based Mic-Mac model is able to reproduce quite accurately the HFB results when 
the liquid-drop parameters of our model have been fitted to the HFB energies. This can be 
seen in Figure \ref{micmac_hfb} where the predictions of our Mic-Mac model in this case are 
given by cyan  symbols.
The Gogny-based WK Mic-Mac model fitting the macroscopic part to 
the full quantal HFB energies reproduces the experimental masses of the selected set of nuclei  
with a similar {\it rms} deviation $\sim$4 MeV as the one provided by HFB calculation, pointing out that the Mic-Mac model 
is a consistent approach and captures the essential physics of the full quantal calculation.

 Our results  show that the energies calculated  using our WK Mic-Mac 
model  with the macroscopic part fitted to the experimental data are to some extent
independent of the external potential used to determine the microscopic part. This is due 
to the fact that the relatively small differences in the microscopic energies computed with
different external potentials can be easily absorbed by the large macroscopic part through 
a variation of the liquid drop parameters.
In this respect, it is expected that the energies predicted by our Gogny-based model
starting from a different Gogny interaction, say D1M \cite{goriely09} for example, would predict on average similar 
energies if the parameters of the macroscopic part are fitted to the experimental data, 
the differences with the results reported in this work, obtained using 
the D1S force, being relatively marginal. 

It is well known that the D1S Gogny interaction suffers from a drift in the energy with respect
to the experimental values when the number of neutrons increases for a given nucleus (see \cite{pillet17} and
references therein) as can be clearly seen in Figure \ref{micmac_hfb}.
To overcome this deficiency, new Gogny interactions of the D1 family, namely D1N \cite{chappert08}
and D1M \cite{goriely09} were proposed. These interactions include in the fitting protocol of 
their parameters new constraints such as, among others, that of reproducing qualitatively the trends of a 
microscopic equation of state in neutron matter. As can be seen in Figure~\ref{micmac_hfb}, 
the Mic-Mac model based on the Gogny D1S force also removes the drift in the binding
energies along isotopic chains. Therefore it is expected that this Mic-Mac model can reproduce the
experimental energies in heavy neutron rich nuclei better than the HFB calculations using in both cases the
same D1S interaction. To analyze more in detail the differences between the full HFB and  the Mic-Mac 
ground-state energies, we report in Table \ref{hfb_micmac_Pb} the binding energies along the Pb isotopic 
chain computed at HFB and Mic-Mac levels. From this Table it is seen that the HFB energies  in this 
isotopic chain exhibit a systematic behaviour with respect  to the Mic-Mac Gogny results obtained
 in the present work. In particular,  we see that for neutron-deficient Pb isotopes, both, HFB and 
Mic-Mac,  agree well with the experiment. With increasing neutron number, the two  predictions deviate 
from each other. The Mic-Mac  results remain close to the experiment, but the HFB results deviate strongly from it. 
As the shell closure approaches, the Mic-Mac results  start to deviate from the experiment, whereas 
the HFB  values go on improving. Away from the shell closure, the Mic-Mac  calculations again improve,
whereas HFB starts deviating from experiment ($^{214}$Pb) as a consequence of the energy drift
mentioned before.

\begin{table}[htb]
\caption{Binding energies (in MeV) for Pb isotopes. Those reported at \cite{luis} and the
experimental values are also quoted for comparison. Notice that the contribution 
to the experimental energies from electronic binding has been removed here.}
\begin{center}
\begin{tabular}{c|c|c|c} \hline
 $A$   &  This work   & Ref. \cite{luis} & Expt.     \\ \hline
~~178~~& ~~-1367.60~~ & ~~ -1369.13~~     &~~-1368.40~~ \\
  180  &   -1389.28   &    -1389.94       &  -1390.05  \\
  182  &   -1410.37   &    -1410.26       &  -1411.08  \\
  184  &   -1430.90   &    -1430.09       &  -1431.45  \\
  186  &   -1450.88   &    -1449.44       &  -1451.23  \\
  188  &   -1470.31   &    -1468.37       &  -1470.50  \\
  190  &   -1489.26   &    -1486.88       &  -1489.25  \\
  192  &   -1507.70   &    -1505.02       &  -1507.54  \\
  194  &   -1525.58   &    -1522.79       &  -1525.32  \\
  196  &   -1542.83   &    -1540.20       &  -1542.62  \\ 
  198  &   -1559.33   &    -1557.26       &  -1559.45  \\
  200  &   -1575.19   &    -1573.96       &  -1575.79  \\
  202  &   -1590.56   &    -1590.28       &  -1591.63  \\
  204  &   -1605.48   &    -1606.21       &  -1606.94  \\
  206  &   -1619.91   &    -1621.66       &  -1621.76  \\
  208  &   -1633.29   &    -1636.44       &  -1635.86  \\
  210  &   -1643.09   &    -1643.79       &  -1644.98  \\
  212  &   -1652.12   &    -1650.81       &  -1653.95  \\
  214  &   -1660.87   &    -1657.49       &  -1662.72  \\ \hline
\end{tabular}
\end{center}
\label{hfb_micmac_Pb}
\end{table}

\subsection{Comparison with other Mic-Mac models}

\begin{figure*}[htb]
\centering \includegraphics[scale=0.4]{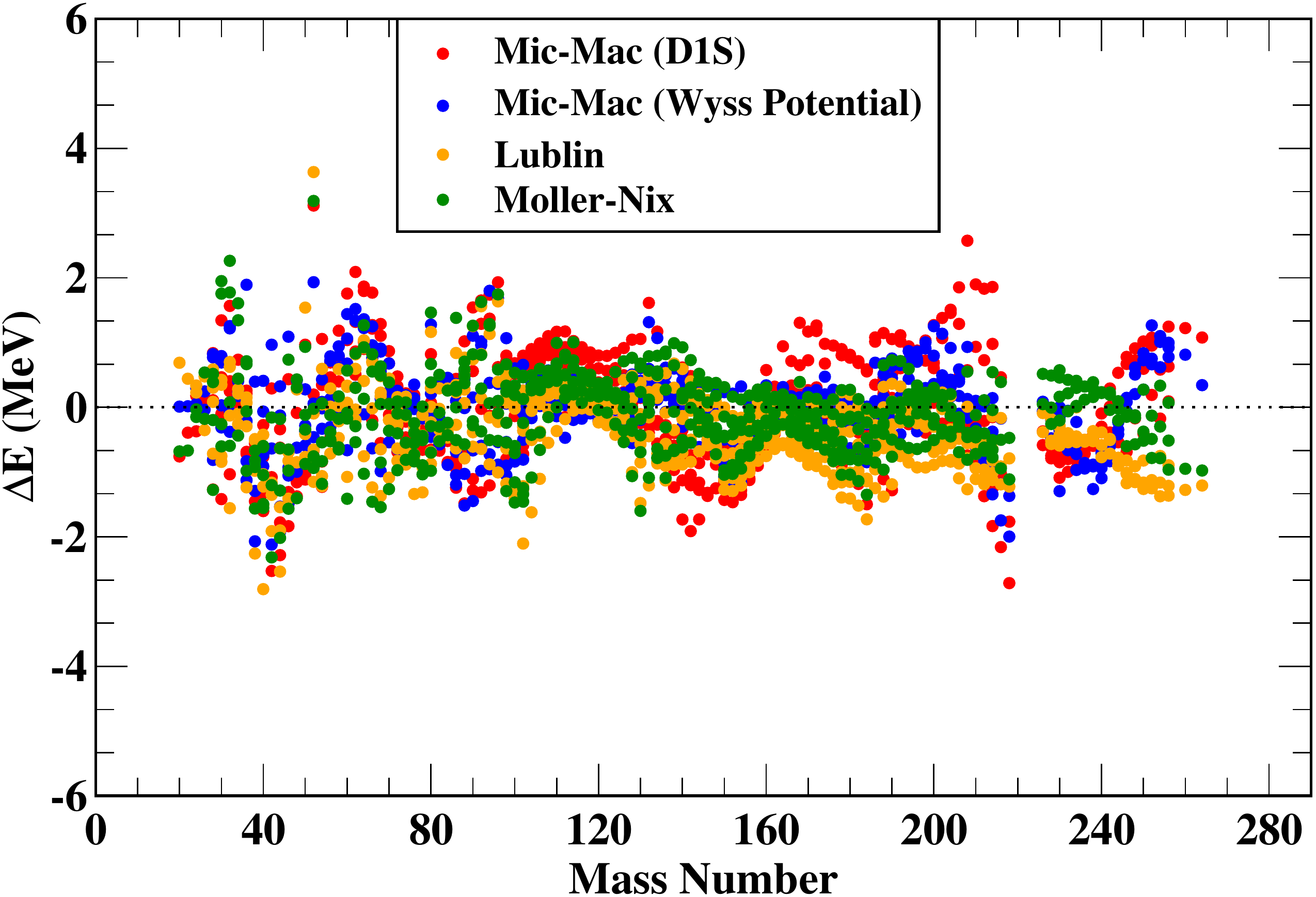}
\caption{Residues with respect to experiment for the energies of 551 spherical and deformed even-even nuclei
obtained with the Mic-Mac models FRDM \cite{moller97}, LSD \cite{pomorski03} and our WK model
using the Gogny-based and the phenomenological (Wyss) potentials as external mean fields.}
\label{difbeall}
\end{figure*}

\begin{figure}[htb]
\centering \includegraphics[scale=0.4]{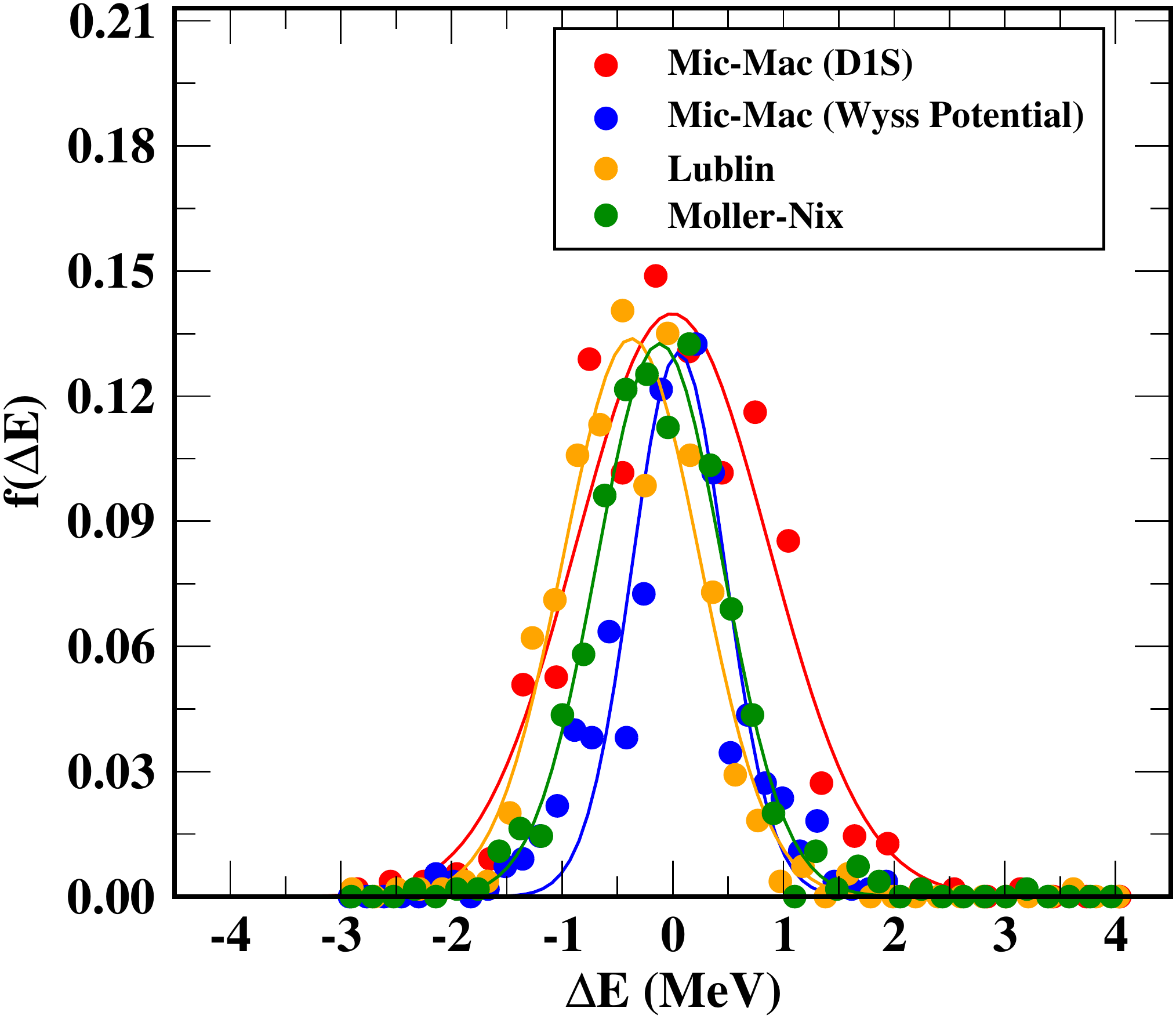}
\caption{Binned data and the corresponding Gaussian fits for the different
Mic-Mac models considered in the present study (see text for more detail).}
\label{statis}
\end{figure}

In this subsection we want to compare the predictions of our Gogny-based Mic-Mac models
with the results provided by very well-known Mic-Mac models such as the FRDM of M\"oller
and Nix \cite{moller97}, the LSD of Pomorski and Dudek \cite{pomorski03}
and the WK Mic-Mac model based on the phenomenological Wyss potential \cite{bhagwat10,bhagwat12}.
 These comparisons are performed for the chosen set of 551 spherical and deformed
nuclei with well-determined masses according to the Audi 2012 evaluation \cite{audi12}.
To this end, we display in Figure \ref{difbeall} the residues with respect to the experimental
energy predicted by the FRDM, LSD, WK Wyss potential and the Gogny-based 
WK Mic-Mac model developed in this work. From Figure \ref{difbeall} we can see that, globally, 
all the considered Mic-Mac models are quite equivalent for describing ground-state energies
with residuals that are not larger than $\pm$ 2 MeV along the whole periodic table.      
All the considered models show, globally, similar trends, with the largest residues corresponding 
to magic numbers. Another common property of these residues is the fact that they are 
relatively larger for low mass than for heavy mass nuclei. 
The fact that all these models qualitatively behave more or less alike needs a more detailed analysis.
In order to do so, we have binned the residues $\Delta E = E_{cal.} - E_{expt}$, i.e. the
difference between the calculated and experimental energies, in a 
suitable way to get the normalized frequency distribution. The bin size was chosen carefully 
through the well-known Freedman-Diaconis procedure \cite{freedman81,izenman91}. 
It is well known that this choice
of the bin size is quite robust, and works well for a range of underlying probability
distributions, so long as the probability distributions are square integrable
functions.

\begin{figure*}[htb]
\centering{\hbox {\includegraphics[scale=0.40]{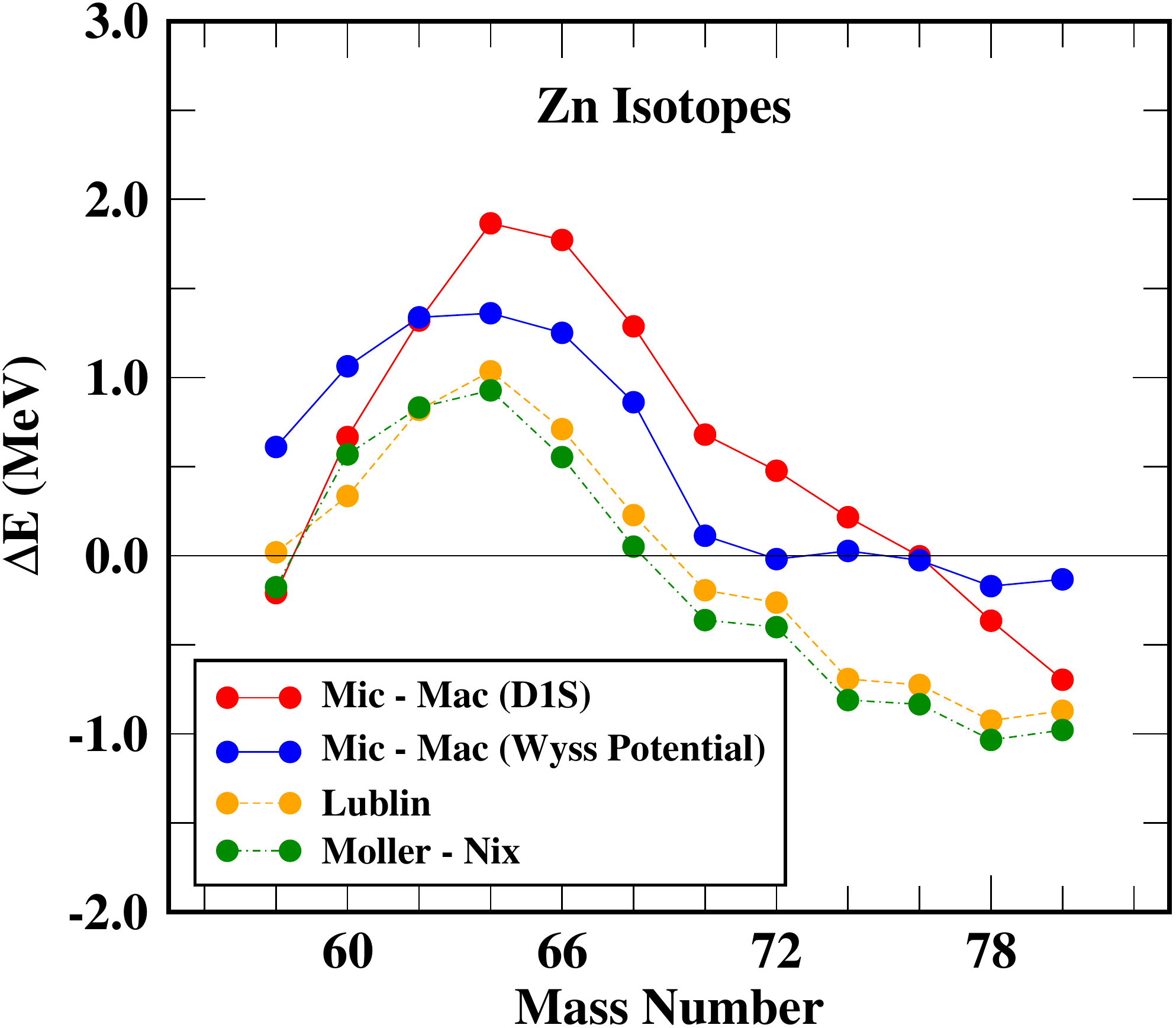} 
\hspace{\fill}
                  \includegraphics[scale=0.40]{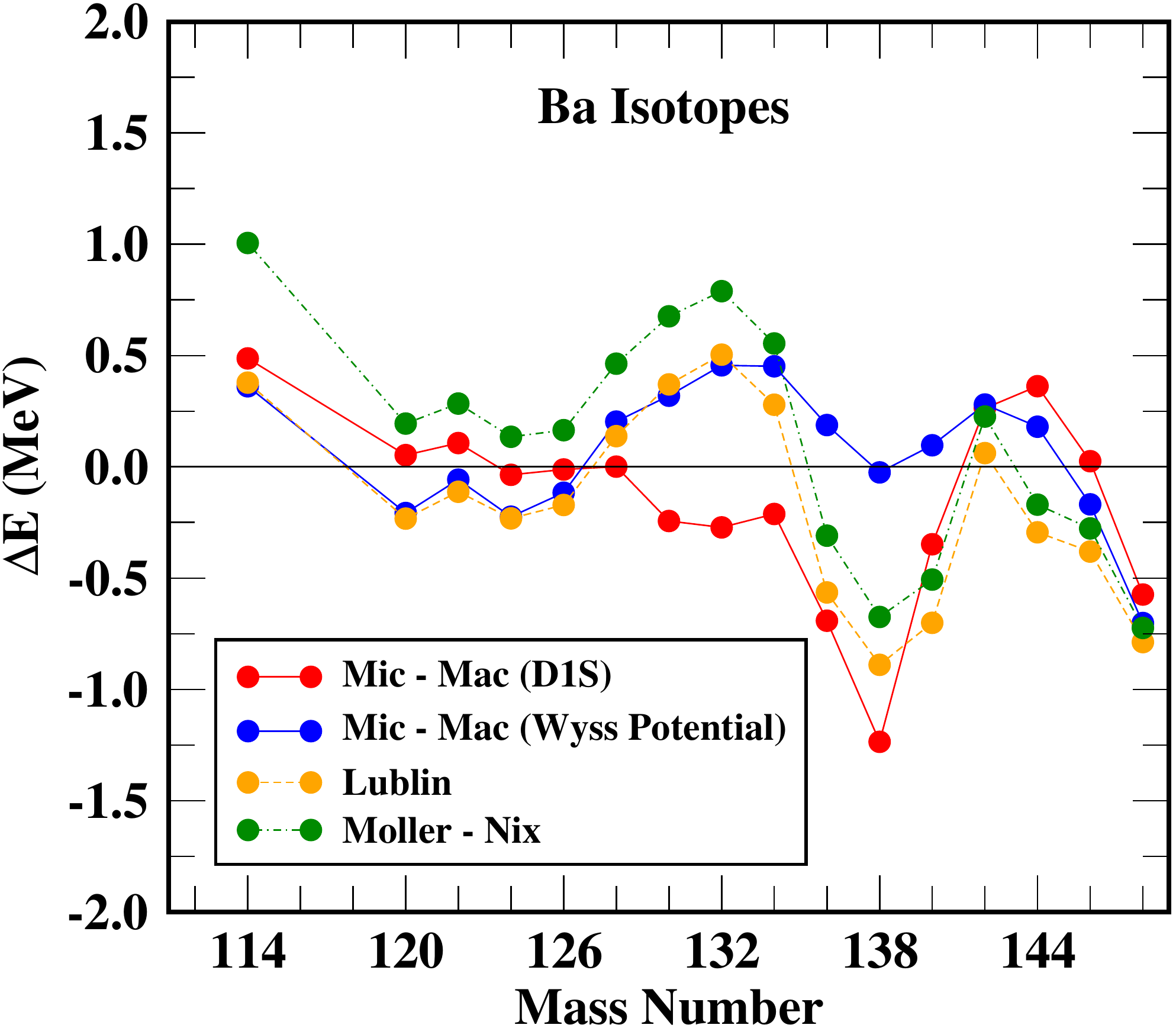}}}
\centering{\hbox {\includegraphics[scale=0.40]{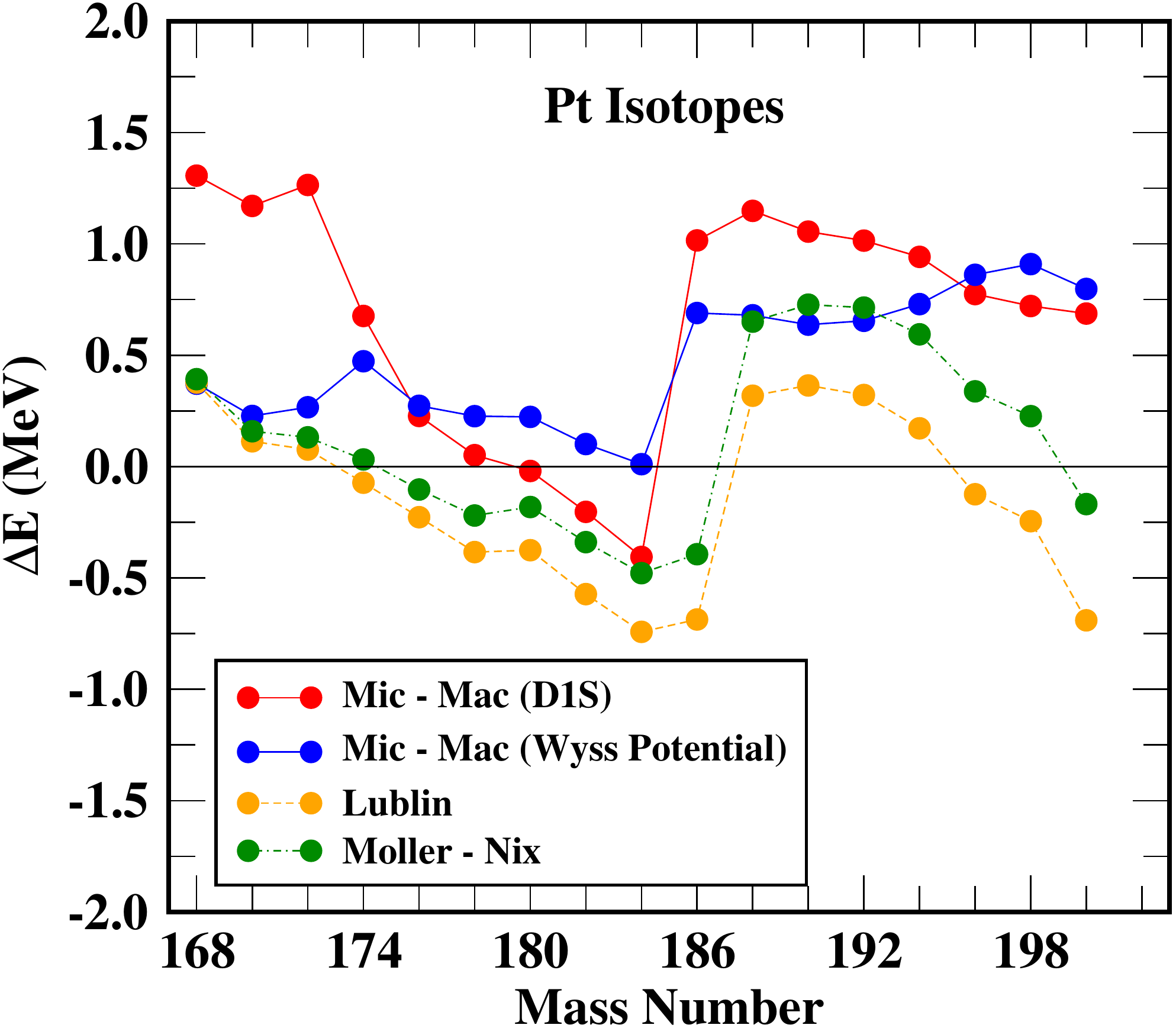}
\hspace{\fill}
                  \includegraphics[scale=0.40]{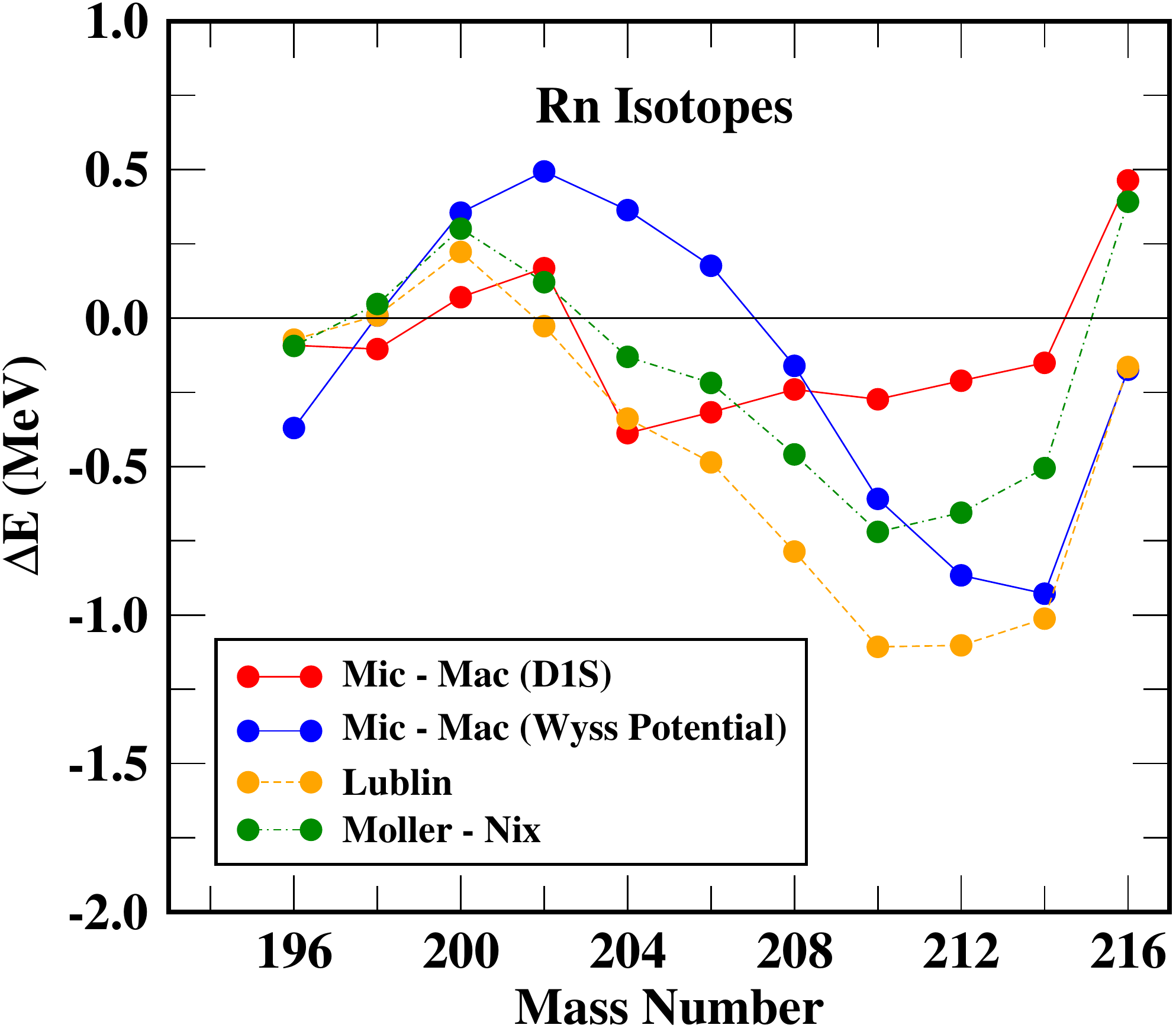}}}
\caption{The difference between the calculated and the experimental
\cite{audi12} energies for Zn, Ba, Pt and Rn isotopes.}
\label{zn-pt}
\end{figure*}

\begin{figure*}[htb]
\centerline{\hbox {\includegraphics[scale=0.40]{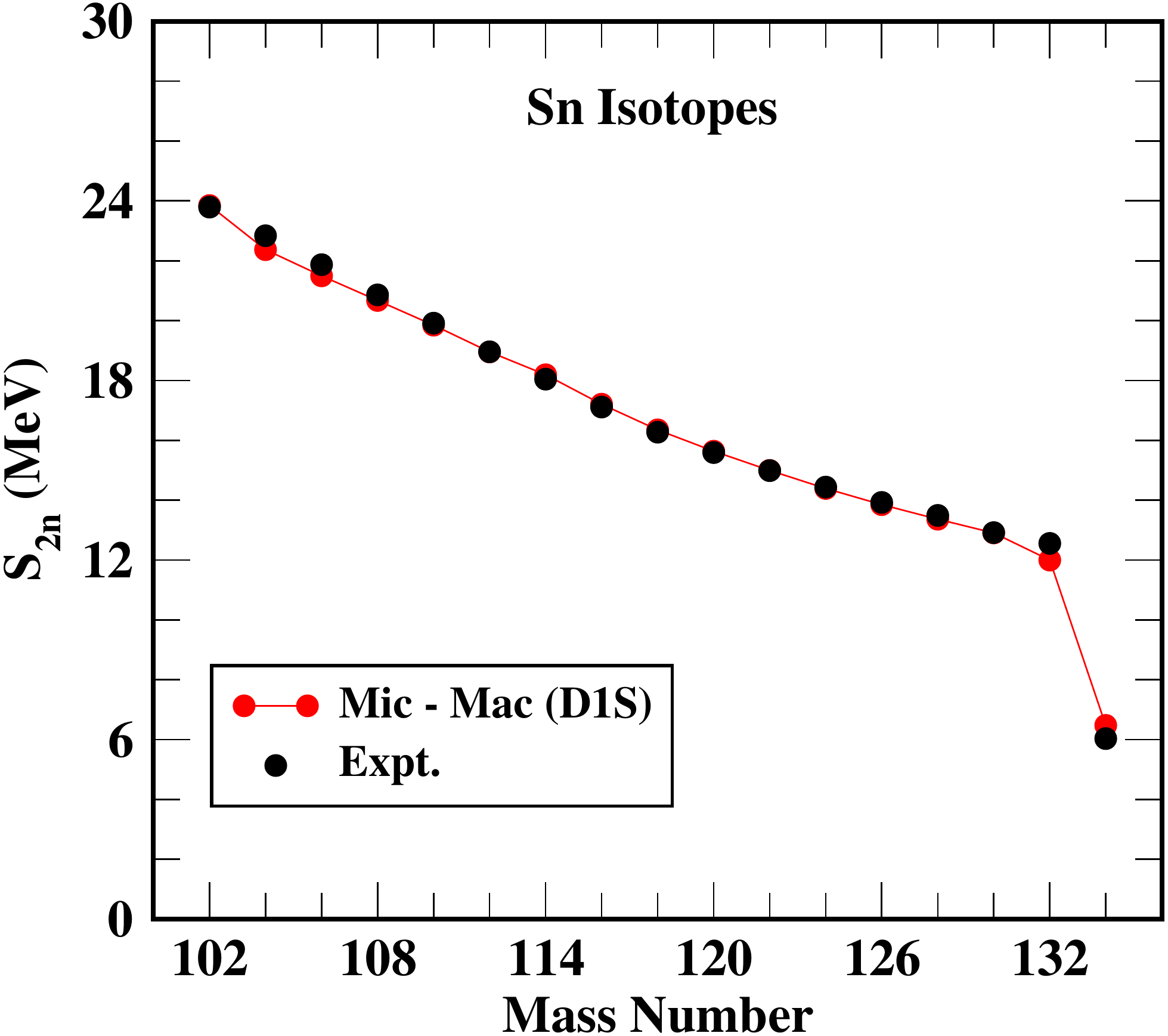} 
\hspace{\fill}
                  {\includegraphics[scale=0.40]{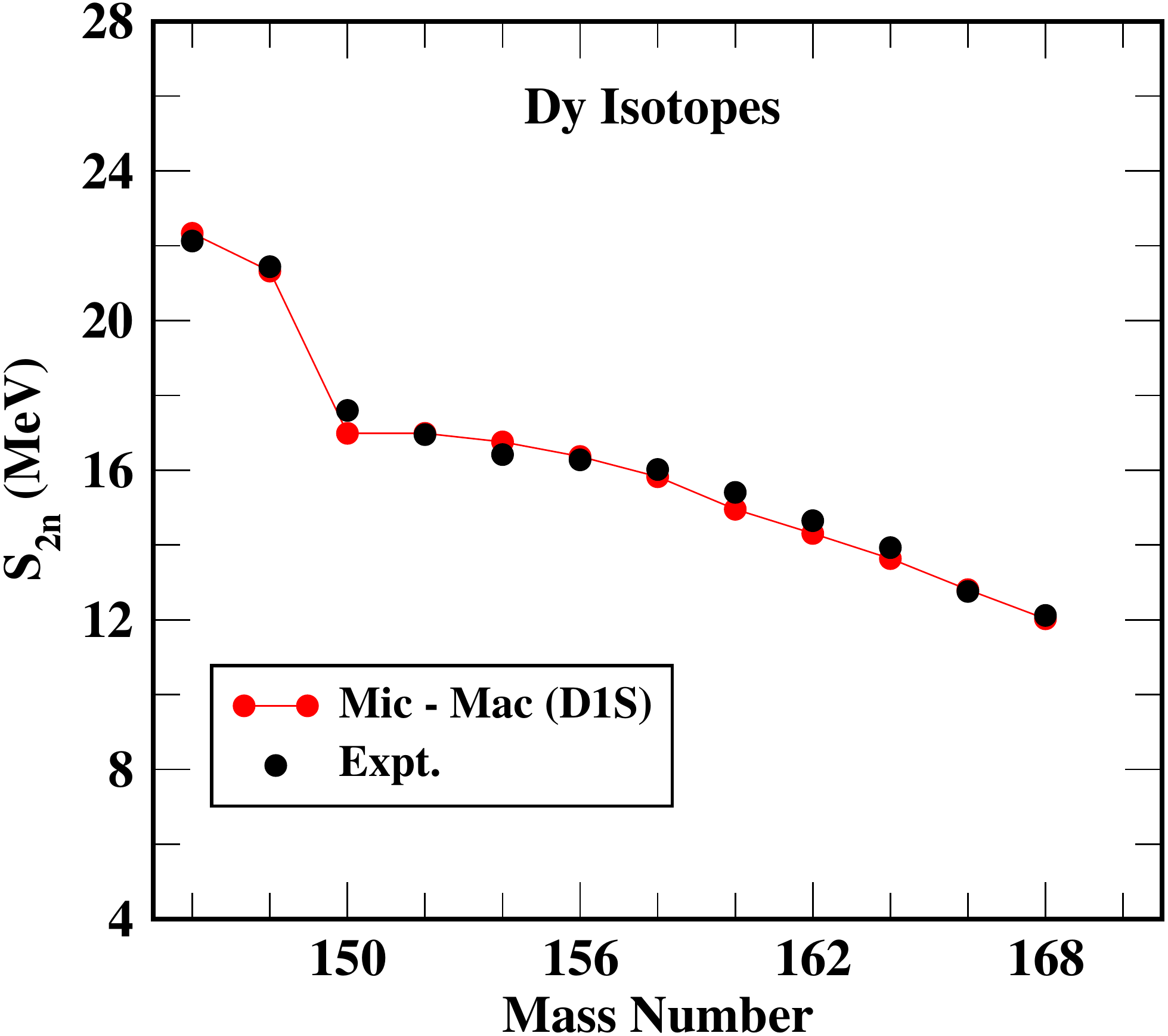}}}}
\centerline{\includegraphics[scale=0.40]{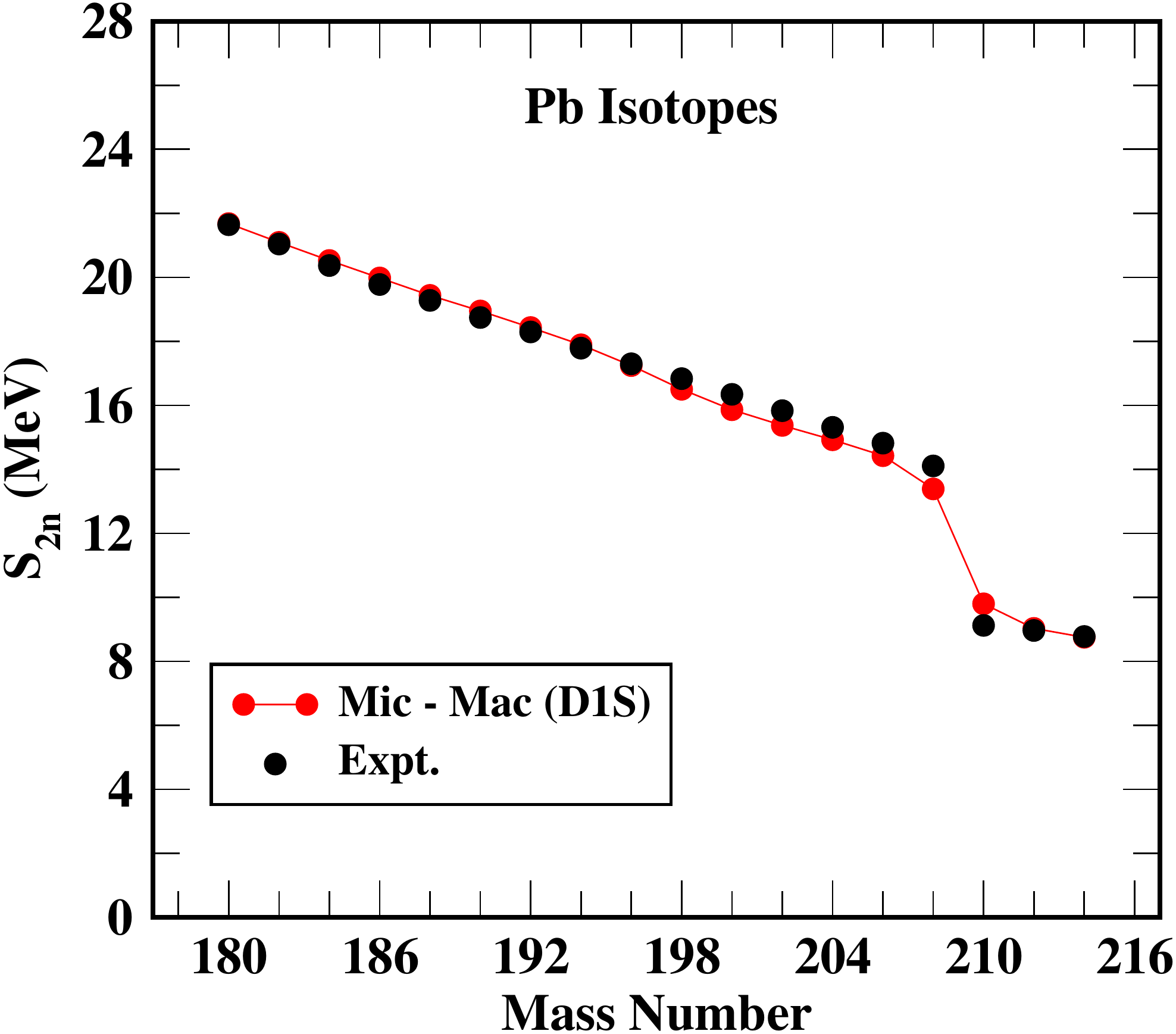}}
\caption{Calculated and experimental values of two-neutron separation energies.}
\label{sep1}
\end{figure*}

The binned data plotted along with the corresponding fitted Gaussian profiles
are displayed in Figure~\ref{statis}. We can see that all the four sets of data 
yield almost Gaussian profiles, with correlation coefficients greater than 0.95 
in all the four cases. All the distributions have a central peak of height $\sim$0.13 
at $\Delta E \sim 0$, indicating that about 13\% of the data is described with deviation 
of $\sim 0$ with respect to the experimental data. Apart from the different standard 
deviations in all the four models, the profiles of the residues were found to be very similar, 
supporting the previous observation that all the four mass models are more or less equivalent,
globally speaking.
A more detailed inspection of this Figure~\ref{statis} shows that our WK models are well centered 
around $\Delta E \sim 0$, while LSD and FRDM show a small shift towards negative $\Delta E$ values, 
which is more important for the LSD data. 
This fact indicates that on average our WK results
are well scattered around the experimental energies while LSD and FRDM show a slight tendency of
 overbinding, at least for the considered set of nuclei. The widths of the Gaussian fits 
suggest that, for the considered set of nuclei, the quality of the WK results using the phenomenological 
Wyss potential is somewhat better than the quality of the predictions of the FRDM and LSD calculations.
From this Figure it is also clear that width of the Gaussian associated to the WK Gogny-based potential
calculation is larger than the other widths displayed in the Figure, pointing out that the energy 
description for the set of considered nuclei provided by the WK Gogny-based Mic-Mac model is a fringe
worse than the one obtained using the other models considered in this analysis.

In order to investigate the predictive power of the WK Gogny-based model
in different regions of the nuclear chart, we show in Figure \ref{zn-pt} the residues with 
respect to the experimental values of the energies along the Zn, Ba, Pt and Rn
isotopic chains computed with our Gogny-based WK model in comparison with the 
predictions of the FRDM, LSD and  WK Wyss potential.
For Zn isotopes, the FRDM and LSD predictions are somewhat better than the  
ones of both WK models for mass numbers between $A$=60 and $A$=70, and the opposite is 
true for the heaviest isotopes of the chain, where the mass  numbers are in the range between
$A$=74 and $A$=80. However, in general, the predictions of the WK Gogny-based model
 underbinds the experimental energies, mainly around $A$=60--66.
For Ba isotopes, the predictions of the FRDM, LSD and   both WK models agree 
reasonably well except, may be, around the magic neutron number $N$=82 (corresponding to 
$^{\rm 138}$Ba) where the 
 Gogny-based WK calculation predicts larger differences with respect to the experiment than
the FRDM, LSD  and WK Wyss models. For Pt isotopes, the FRDM, LSD and both WK residues 
show, qualitatively, a similar behaviour. However, in general, the residues corresponding 
to the WK Gogny-based model are larger than the ones predicted by the FRDM, LSD and WK Wyss calculations.
For Rn isotopes, the WK Gogny-based model predicts similar binding energies than the other Mic-Mac 
models considered in the range of mass numbers between $A$=196 and $A$=206, while in the 
intermediate region with mass numbers between $A$=208 and $A$=210,
the residues predicted by the WK Gogny-based model are a bit smaller than the ones 
found in the  FRDM, LSD and WK Wyss calculations. Let us mention that the good agreement between the 
residues obtained in the FRDM and LSD calculations is, actually, due to the fact that both
models use the same microscopic part. 

A more quantitative information about the goodness
of the different Mic-Mac  models analyzed in this work is provided by the energy {\it rms} deviations, 
which for the set of considered nuclei are 635 keV (FRDM), 731 keV (LSD), 609 keV WK(Wyss) 
and 834 keV WK(Gogny). The fact that the {\it rms} deviation 
predicted by our WK Gogny-based calculation (834 keV) is larger than the one obtained with our WK method using
the phenomenological Wyss potential (609 keV), which in turn is similar to the {\it rms} deviation corresponding 
to FRDM and LSD models for the same set of nuclei, can be appreciated in Figure~\ref{difbe}. In this
figure we can see that the prediction of the WK model based on the Wyss potential gives a better 
description of the experimental energies in the range between $A \sim$100 and $A \sim$200 than the 
WK Gogny-based calculation. 
These facts show the limits of the predictive power of the Mic-Mac model based
on the D1S interaction and suggest the two following comments. On the one hand, the use of a more accurate
Gogny interaction for describing finite nuclei, as for example D1M \cite{goriely09}, 
which is free of the energy drift discussed before, may slightly improve the quality of the description of ground-state 
energies with the WK Gogny-based model.
Although the global quality of the WK Gogny-based model is lower by a very small amount for describing
ground-state energies as compared with the predictions  of the other Mic-Mac models considered in this
work, it is accurate enough to give predictions in good agreement with the experimental data.
As an example, we display the two-neutron separation energies for Sn, Dy and Pb isotopes in Figure~\ref{sep1}. The excellent
agreement between calculations and experiment can be clearly seen in the figure.
However, despite the globally good performance, we must admit that the use of a more microscopically based mean field as the one used here, 
which is based on the Gogny D1S force, does not show any decisive advantage over, e.g, the purely phenomenological Wyss potential,
in what concerns the calculation of nuclear masses.  

\begin{figure*}[htb]
\centering \includegraphics[scale=0.4]{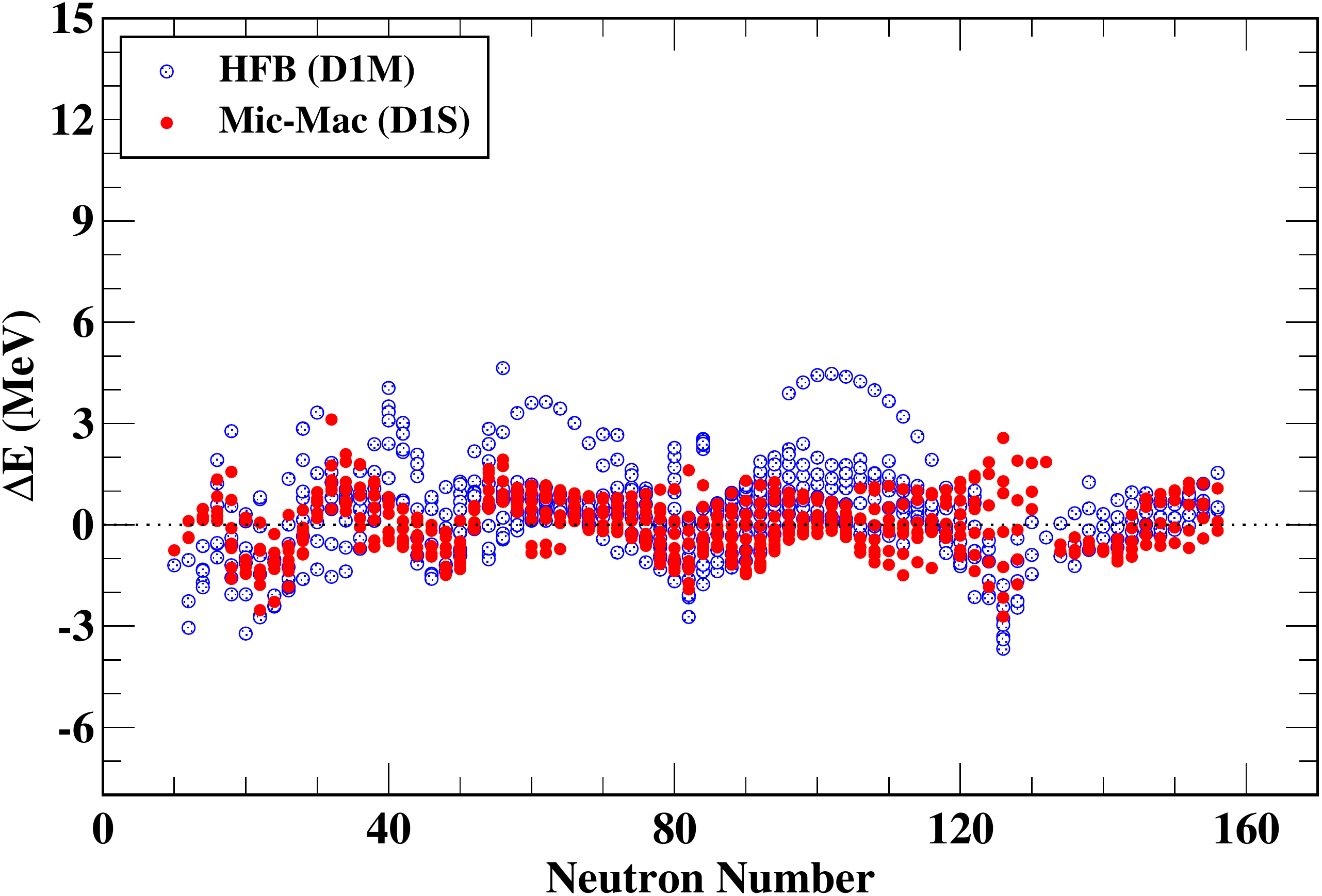}
\caption{Residues with respect to experiment for the energies of 551 spherical and deformed even-even nuclei
calculated with the WK Mic-Mac model based on the Gogny D1S force (filled red symbols)
and the HFB results calculated with the Gogny D1M force (open blue symbols).}
\label{difbed1m}
\end{figure*}

However, some remarks on the underlying physics of our Mic-Mac approach are in order.  The Gogny D1S Mic-Mac results clearly show 
that the relatively large {\it rms} deviation of 3.95 MeV (see Table \ref{Table1}) of the microscopic HFB D1S masses is not necessarily 
to be incriminated to a bad behavior of 
the shell structure (given by the shell correction energies) but rather due to a deficiency of the bulk behavior. It is most obvious 
that the LDM part of our Mic-Mac model saves the bulk part from the neutron drift. But even if we take a Gogny version which is 
free from the drift, as for instance D1M, for which the {\it rms} deviation is 1.34 MeV, our Mic-Mac model for the considered nuclei still 
performs essentially better (as a reminder the {\it rms} deviation is 0.834 MeV in our model), as it can also be seen seen from Figure 
\ref{difbed1m} where the residuals obtained from the D1M HFB calculation and from our Mic-Mac model are plotted as a function of the 
neutron number. This suggests that, in general, a HFB calculation with effective Gogny forces cannot reproduce at the same time 
good bulk and good shell effects behavior. This fact is clear for the calculation with the D1S interaction, but can also be appreciated 
in the case of the D1M force as just explained. Actually it can be concluded that it is the bulk part of the HFB energy computed with 
the Gogny force which is deficient and can 
be improved by the LDM contribution. For example the difficulty for evaluating the zero point motion contribution is automatically included 
in the LDM part. This conclusion is in fact quite general and it arises from the lack of flexibility of the effective 
interaction, where both microscopic and macroscopic parts are coupled. It is by means of a highly sophisticated 
functional with 30 parameters including an additional Mac part (Wigner term) that the semi-microscopic HFB model of the Brussels group 
\cite{goriely13} yields simultaneously good bulk and good shell structure. In our opinion, there is a room for for further improvement 
in the Mic-Macmodels. 
This concerns not so much the LDM part, which we think is determined in a quite optimal way, as the shell correction part
with, for instance, its behaviour around magic and doubly magic nuclei.
It is one of our objectives to work on this in the future.

 \begin{figure*}[htb]
 \centering {\hbox {\includegraphics[scale=0.7]{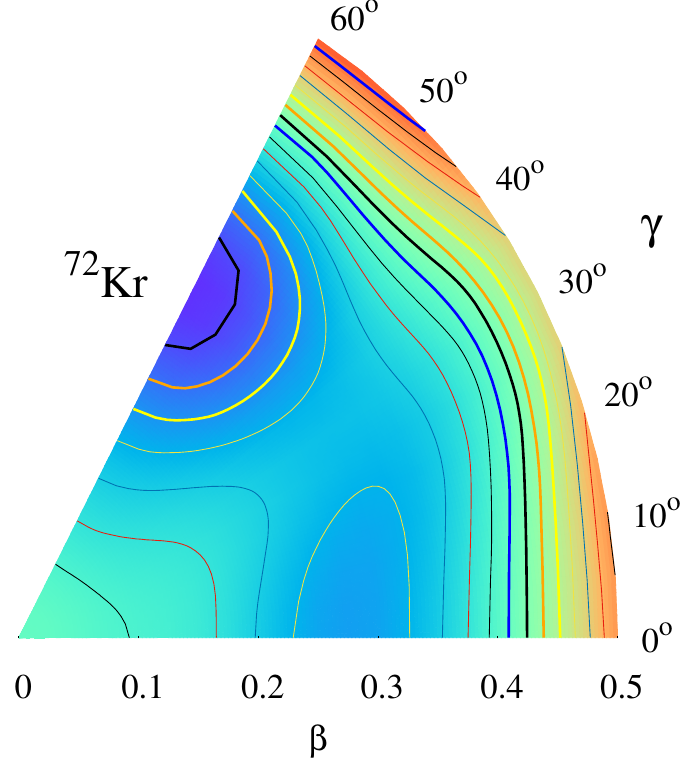} \quad
                    \includegraphics[scale=0.7]{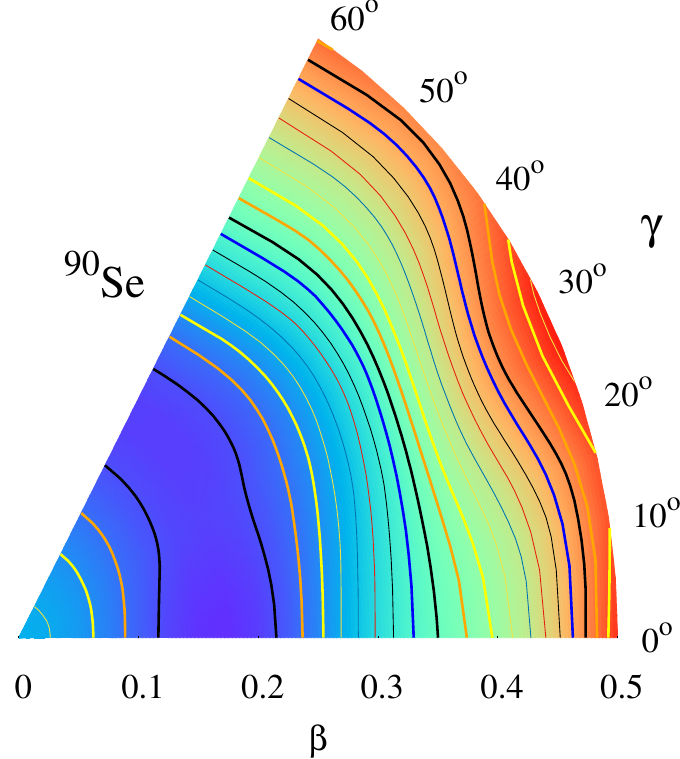} \quad
                    \includegraphics[scale=0.7]{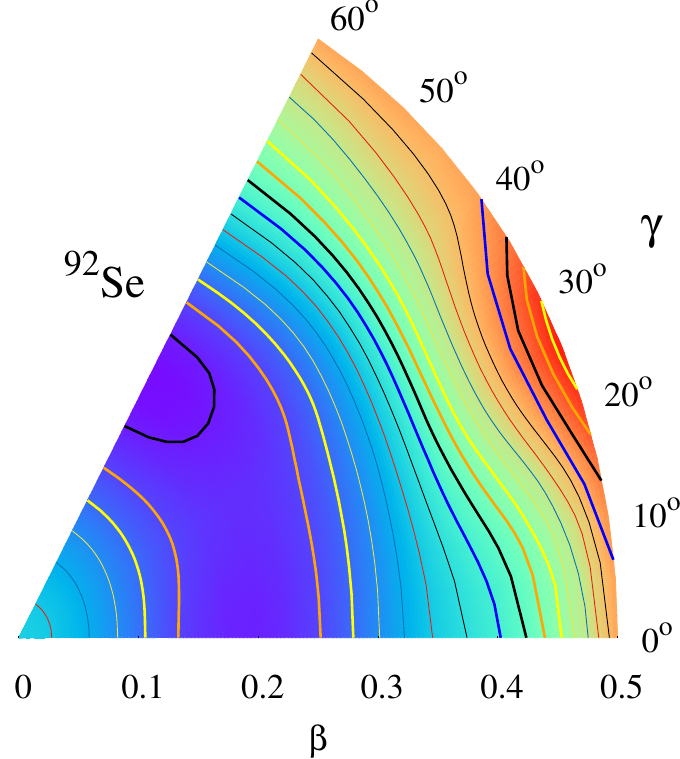}}}
 \centering {\hbox {\includegraphics[scale=0.7]{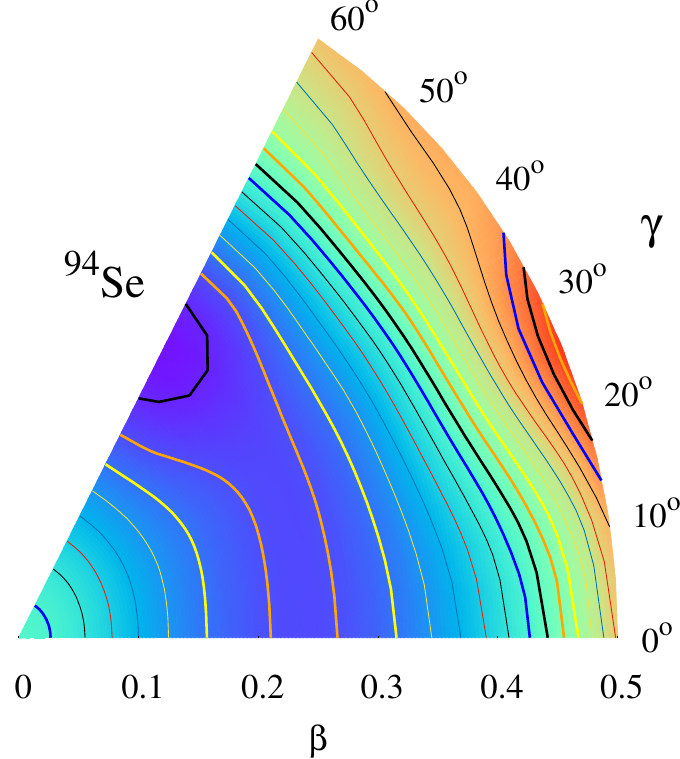} \quad
                    \includegraphics[scale=0.7]{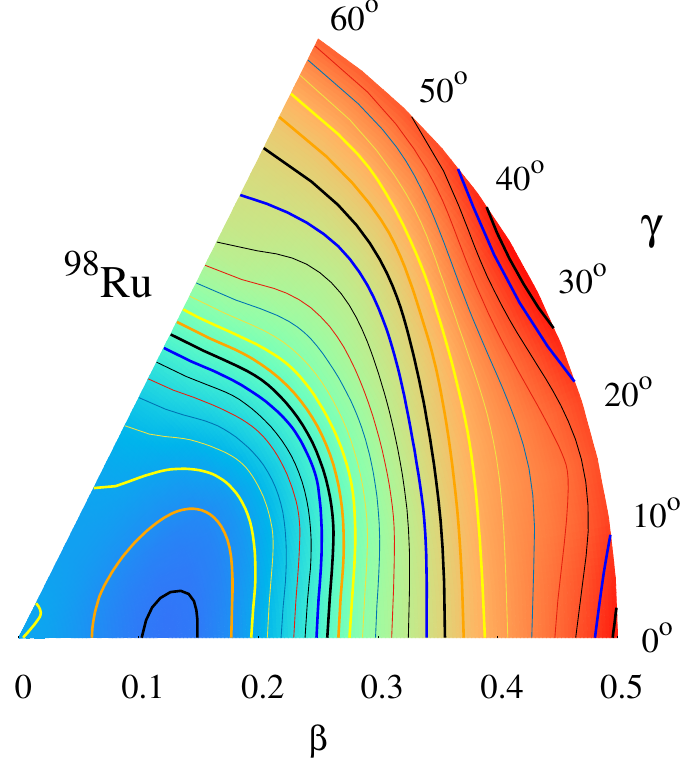} \quad
                    \includegraphics[scale=0.7]{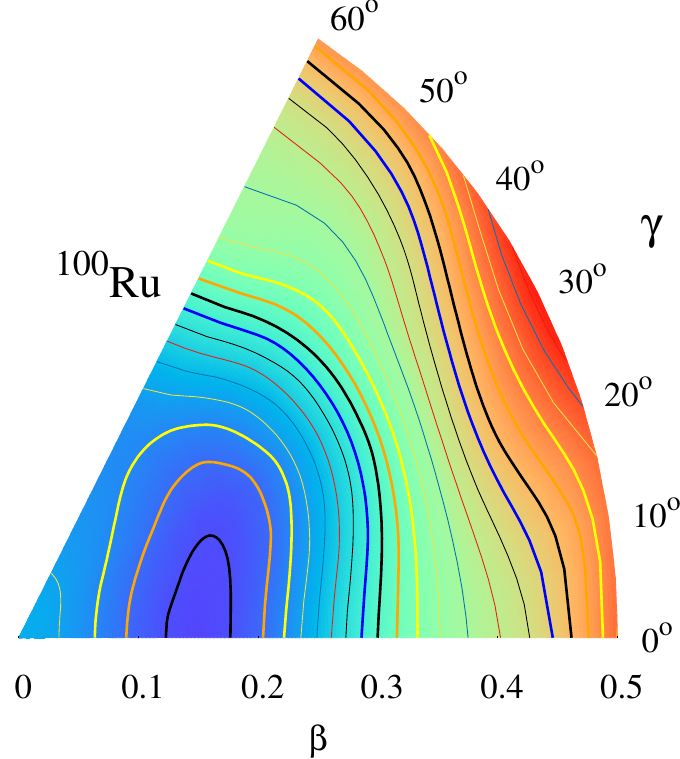}}}
 \centering {\hbox {\includegraphics[scale=0.7]{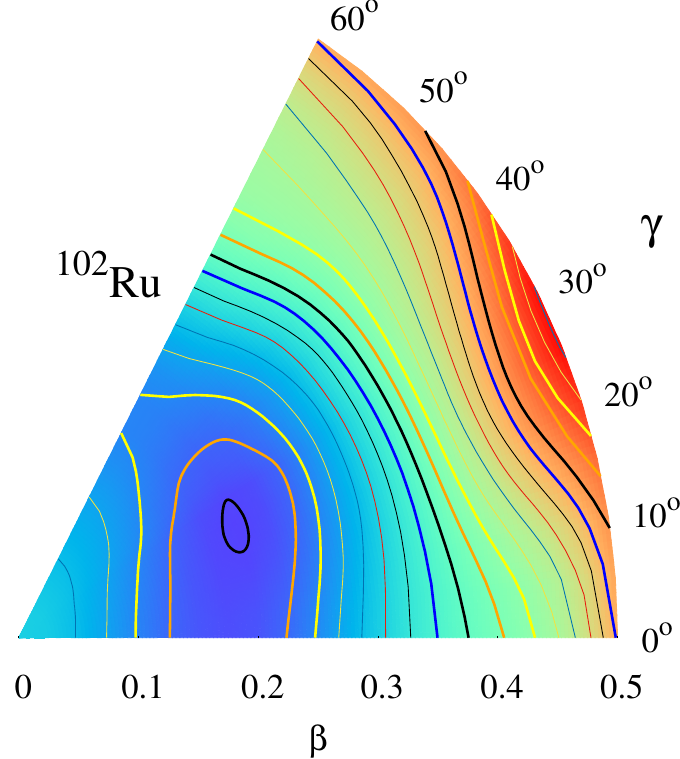} \quad
                    \includegraphics[scale=0.7]{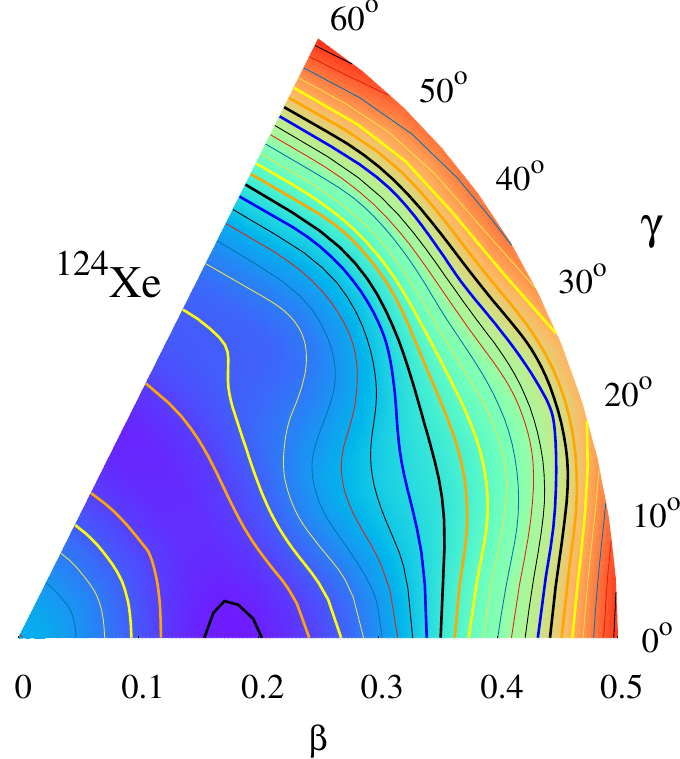} \quad
                    \includegraphics[scale=0.7]{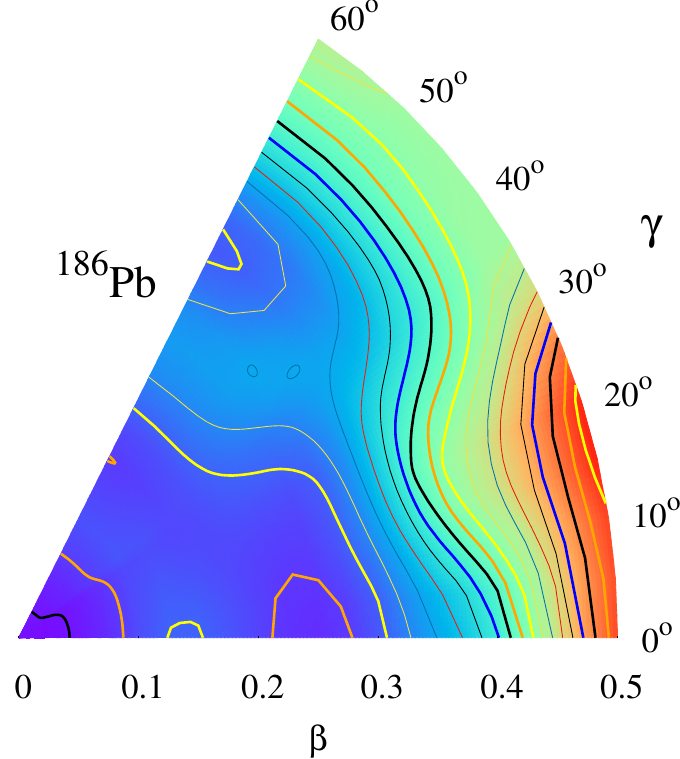}}}
\caption{Potential energy surfaces for a set of nuclei spanning almost the entire
periodic table. See text for details.}
\end{figure*}

\section{Potential Energy Surfaces}

The potential energy surfaces (PES) have been generated by suitably transforming
the binding energies into Cartesian representation. Considering the fact that
 the maxima and the minima differ by at the most 10 MeV in our present 
set of nuclei, the sampling has been done with a bin size of 0.4 MeV. 

As representative cases, we plot the potential energy surfaces for $^{72}$Kr, 
$^{90,92,94}$Se, $^{98,100,102}$Ru, $^{124}$Xe and $^{186}$Pb.
These have been so chosen as to demonstrate existence of well-defined 
minima (prolate or oblate), the shape-coexistence phenomenon, as well as $\gamma$-softness.

As expected, $^{72}$Kr turns out to be very well defined oblate, with deformation 
parameter ($\beta_2$) of the order of 0.3. This is in agreement with the other 
Mic-Mac calculations, such as M\"oller-Nix and our calculation with Wyss potential. 
The $^{72}$Kr nucleus turns out to have a second prolate minimum at $\beta_2 \sim 0.275$
around 1.5 MeV away from the deepest minimum. This topography is at variance with that 
predicted by the pure HFB calculation with the D1S force, which predicts the existence of two closely 
spaced minima in the oblate region \cite{hilaire08}. 

The three Selenium isotopes exhibit the shape coexistence phenomenon. In particular, 
 the ground state of the nucleus $^{90}$Se is predicted to be prolate 
with $\beta_2 \sim 0.175$, possesses 
a second minimum for oblate $\beta_2 \sim 0.2$, with a difference between the 
two of the order of just 200 keV. On the other hand, $^{92,94}$Se are predicted
to have oblate minima, with the second minima (prolate) which are $\sim$200 keV 
and $\sim$600 keV away respectively. Further, the potential energy surface 
as a function of $\gamma$ for $\beta_2$ around 0.2 turns out to be quite flat, 
particularly for $^{90}$Se, indicating a possible existence of $\gamma$-softness.
These observations are similar to that reported in Ref.~\cite{nomura17}.
Out of the three Ruthenium isotopes, $^{98,100}$Ru are strongly prolate, with a rather flat
PES for $\gamma$ up to 20$^\circ$. The nucleus $^{102}$Ru turns out to be $\gamma$-soft, with a
triaxial minimum appearing for $\beta_2 \sim 0.2$ and $\gamma \sim 20^\circ$, 
which is in agreement with the M\"oller-Nix predictions. 

The nucleus $^{124}$Xe possesses a prolate minimum ($\beta_2 \sim 0.2$) and 
the second minimum is oblate  which is about 400 keV away from the deepest 
minimum. Interestingly, the PES for this nucleus turns out to be quite flat as a
function of $\gamma$, for $\beta_2$ values around 0.2, hinting towards a possible
existence of $\gamma$-softness. Notice that this behaviour is similar to that 
observed for Selenium isotopes.  
The nucleus $^{186}$Pb
turns out to have a very rich structure in its PES. The minimum in this case
is spherical as expected, but there are also several possible deformed solutions 
within $\sim$ 400 keV of the lowest solution. From the plot of the PES it can be 
seen that there is a well-defined prolate minimum at $\beta_2 \sim$ 0.25 and two rather 
well-defined oblate minima, one at $\beta_2 \sim$ 0.30 and the other one at $\beta_2 \sim $ 0.10,
which is located in a very small contour line along the oblate axis. This pattern for the 
PES of the nucleus $^{186}$Pb is in agreement with the results reported by M\"oller 
{\it et al.} \cite{moller2008}.

\section{Summary and Conclusions}

 In this paper we used mean field potentials which have been derived via the semiclassical ETF method, 
including $\hbar^2$ corrections, from the Gogny-HF approach using the D1S force. The semiclassical mean-fields 
 are local with the effective mass incorporated into the potentials \cite{soubbotin00}. This choice was dictated by the numerics. Using those mean-field 
potentials in a one-shot diagonalization (the so-called expectation-value method \cite{brack85,bohigas76}) allows one
to recover very accurately the fully microscopic results (see Paper I \cite{bhagwat20}). 
In this work, we therefore tried to combine the Mic-Mac approach and the 
microscopic HFB method with the Gogny force. We, thus, have firstly taken the Liquid Drop Model for the smooth part of 
the energy, whose parameters are fitted to the experimental masses, plus second  the shell correction part obtained with 
our semiclassical Gogny-adopted mean field potentials. This then constitutes a Mic-Mac approach where only the 
shell corrections are directly connected to the parameters of the Gogny force (D1S in the 
present case). Not astonishingly the neutron drift of the binding energies inherent to HFB calculations with the 
D1S force has been eliminated while the shell effects are reproduced very accurately. The Mic-Mac calculations 
performed in this way are found to yield a reasonably good description of ground-state binding energies for the 
nuclei spanning the entire periodic table. The rms-value for binding energies obtained in this way is 834 keV. This 
is a value only very little worse than the ones obtained with the completely phenomenological mean field potential 
of, e.g., Wyss \cite{bhagwat10} 
for the shell effects. The present Mic-Mac calculations tend to perform well in the regions away from shell closures, 
whereas the HFB-Gogny results are found to be better near the shell closures. One of the main conclusions of this work 
is that phenomenological effective forces, like the Gogny or Skyrme forces with a limited number of adjustable 
parameters (around 10) are {\it not} able to yield optimal values for the macroscopic part of the energy and, at the same time,
 the shell model contribution. 
On the contrary, we think that a direct fit of the macroscopic part, via a  Liquid Drop model, can optimise the smoothly 
varying macroscopic part of the energies bringing it very close to its exact value. This can also be seen in Fig.~\ref{difbed1m}
where the average of the remaining fluctuations practically adds up to zero. For example, in this way the average of the zero poin
the smoothly varying macroscopic part of the energies. For example in this way the zero point fluctuations are taken care of 
automatically while in a pure HFB approach they are difficult to calculate and in most HFB approaches they are not determined 
unambiguously. The shell effects can then be added additionally and their theoretical evaluation becomes a separate treatment making 
the whole approach more flexible. Since the shell energies are obtained from the difference between a semiclassical and a quantal 
calculation, errors may cancel out. Nonetheless, there may be more room for improvements in the shell corrections than in the LDM part. 
For example, it has been conjectured that the spin-orbit potential may be responsible for persisting oscillations in the 
differences between theoretical and experimental values. Further investigations along these lines are in 
progress. Let us also point out that our Mic-Mac method based on a effective 
two-body force performs practically as well as the most efficient Mic-Mac models on the market in order to describe 
ground-state properties. Only the mean field HFB approach by the Brussels group can compete with this (actually with a 
slightly better result) at the price to adjust around 30 parameters and still employ a macroscopic piece (the Wigner energy) 
on top of it. Finally, let us mention that the results reported in this work concern ground-state energies only and in order to apply 
this Mic-Mac method to other scenarios where large deformations are needed, such as the description of fission phenomena, 
would require  to modify the distance function used here. Work in this direction is in progress.  

\vskip\baselineskip 
\appendix{{\bf Appendix 1: Order by Order Contributions to WK Energy}}
\vskip\baselineskip 
The WK particle number $N$ and energy can be calculated directly by
Laplace inversion as
\begin{eqnarray}
N~=~{\cal{L}}^{-1}_{\lambda} \left( \frac {Z_{\rm WK}^{(4)}(\beta)} {\beta}
\right)
\end{eqnarray}
and
\begin{eqnarray}
E_{\rm WK}~=~\lambda N~-~{\cal{L}}^{-1}_{\lambda} \left( \frac
{Z_{\rm WK}^{(4)}(\beta)}
{\beta ^2} \right) ,
\label{E_qm1}
\end{eqnarray}
where $\lambda$ is the chemical potential, determined to ensure the correct
particle number, $Z_{\rm WK}^{(4)}$ is the WK partition function up to $\hbar^4$-order,
and ${\cal{L}}^{-1}_{\lambda}$ denotes the Laplace inversion.
These Laplace inversions can be performed analytically.

Let $g_{\rm WK}$ be semi-classical level density (up to $\hbar^{4}$ order, in the 
present context), and $\lambda$ be the chemical potential as defined above. In terms
of these quantities, particle number can be expressed as
\begin{eqnarray} 
N~=~\int_{0}^{\lambda}\,g_{WK}\left(\epsilon\right)\,d\epsilon\, .
\end{eqnarray} 
This expression can be though of as a convolution of $g_{WK}$ and the Heaviside step function ($u$).
Since $Z_{\rm WK}^{(4)}$ is the Laplace transform of $g_{\rm WK}$, the Laplace transform
of convolution of $g_{\rm WK}$ and $u$ is $Z_{\rm WK}^{(4)}\left(\beta\right)/\beta$,
which directly yields Eq. (17). \footnote{If $G$ is the Laplace transform of a function $g$, then 
the convolution $u*g$ has the Laplace transform $G(s)/s$, where, $u$ is the Heaviside step
function.}

On the other hand, the averaged energy ($E_{\rm WK}$) can be expressed in terms of 
$g_{\rm WK}$ as:
\begin{eqnarray} 
E_{\rm WK}~=~\int_{0}^{\lambda}\,g_{\rm WK}\left(\epsilon\right)\epsilon\,d\epsilon \, .
\end{eqnarray} 
Integrating by parts one gets:
\begin{eqnarray} 
E_{\rm WK}&=&\lambda N - \int_{0}^{\lambda} \int^{\epsilon} g_{\rm WK}\left(\varepsilon\right)\,d\varepsilon \, 
d\epsilon \, .
\end{eqnarray} 
Notice that the second term on the right hand side of Eq. (21) can be thought of as convolution of $u$
with the convolution of $u$ and $g_{\rm WK}$. Therefore, Laplace transform of this quantity is 
$Z_{\rm WK}^{(4)}\left(\beta\right)/\beta^2$, which automatically leads to Eq. (18).

The explicit expressions of the energy, which are used in Eq.~(\ref{EJEN}) of the main text,
are as follows (see \cite{bhagwat10} for further details):
\begin{widetext}
\begin{eqnarray}
E_{\hbar^{0}}^{\rm CN} &=& \frac{1}{3\pi^2}\left(\frac{2m}{\hbar^2}\right)^{3/2} \int d\vec{r} \left\{
                 \frac{2}{5}\left(\lambda-V\right)^{5/2} \right\} \Theta \left(\lambda-V\right) \\
E_{\hbar^{2}}^{\rm CN} &=& -\frac{1}{24\pi^2}\left(\frac{2m}{\hbar^2}\right)^{1/2} \int d\vec{r} \left\{
                 \left(\lambda-V\right)^{1/2} \nabla^2 V \right\}\Theta \left(\lambda-V\right) \\
E_{\hbar^{4}}^{\rm CN} &=& - \frac{1}{5760\pi^2} \left(\frac{\hbar^2}{2m}\right)^{1/2} \left[
\int d\vec{r} \left(\lambda-V\right)^{-1/2}\left\{ 7\nabla^4 V\right\} 
\Theta \left(\lambda-V\right) \right. \nonumber \\
  && \left. \hspace{3.00cm}
    + \frac{1}{2} \int d\vec{r} \left(\lambda-V\right)^{-3/2} \left\{ 5\left(\nabla^2V\right)^2+\nabla^2\left(\nabla V\right)^2 
\right\} \Theta \left(\lambda-V\right) \right] \\
E_{\hbar^{2}}^{\rm SO} &=&  \frac{\kappa^2}{6\pi^2}\left(\frac{2m}{\hbar^2}\right)^{1/2} \int d\vec{r} 
\left\{ \left(\lambda-V\right)^{3/2} \left(\nabla f\right)^2 \right\}    \Theta \left(\lambda-V\right)  \\
E_{\hbar^{4}}^{\rm SO} &=&
      \frac{1}{48\pi^2}\left(\frac{\hbar^2}{2m}\right)^{1/2} \int d\vec{r} \left(\lambda-V\right)^{1/2}
\left[ \kappa^2 \left\{ \frac{1}{2}\nabla^2 \left( \nabla f\right)^2 - \left(\nabla^2 f\right)^2
         + \nabla f \cdot \nabla \left(\nabla^2 f\right) \right. \right. \nonumber \\
  & & \left.\left. - \frac { \left(\nabla f\right)^2 \nabla^2 V } { 2 \left(\lambda-V\right) } \right\}
      - 2\kappa^3 \left\{ \left( \nabla f\right)^2 \nabla^2 f - \frac{1}{2} \nabla f \cdot \nabla \left(\nabla f\right)^2 \right\}
      +2\kappa^4 \left( \nabla f\right)^4
\right ] \Theta \left(\lambda-V\right) , \nonumber \\
\label{E_WK}
\end{eqnarray}
\end{widetext}
where CN and SO refer to the contributions arising 
from the central and spin-orbit parts of the partition function,
$\kappa$ is the strength of the spin-orbit interaction, and $f$ is the spin-orbit form factor. 
In the present work the potentials appearing in the above expressions as well as the spin-orbit form factor
are of the generalized Woods - Saxon form. The explicit expressions of the derivatives of the generalized 
Woods - Saxon function have been listed in Appendix - 2.

We shall now demonstrate that the semi - classical energy, as written above, reduces 
to the well known Thomas Fermi form. In order to do so, first, notice that
the chemical potential $\lambda$ here is obtained by demanding the correct particle number. The 
particle number, in turn is obtained by integrating the semi-classical density, which
needs to be expanded only to the second order in $\hbar$ so that the energy is correct
up to the fourth order in $\hbar$ (see, for example, \cite{jennings75}). If $\rho_{\rm WK}$ is the semi-classical
density, we can write:
\begin{eqnarray} 
\lambda N &=& \int d\vec{r}\,\left\{\vphantom{\frac{1}{1}} \lambda\rho_{\rm WK}\right\} 
\Theta\left(\lambda - V\right) \nonumber \\
          &=& \int d\vec{r}\,\left\{\vphantom{\frac{1}{1}}  \left(\lambda - V\right)\rho_{\rm WK}\right\}
\Theta\left(\lambda - V\right) \nonumber \\
          &+& \int d\vec{r}\, \left\{ \vphantom{\frac{1}{1}} \rho_{\rm WK}\,V\right\} 
\Theta\left(\lambda - V\right) \nonumber 
\end{eqnarray} 
Up to leading order, $\rho_{\rm WK}$ is given by:
\begin{eqnarray} 
\rho_{\rm WK} &=& \frac{1}{3\pi^2}\left(\frac{2m}{\hbar^2}\right)^{3/2} 
\left(\lambda - V\right)^{3/2}\Theta\left(\lambda - V\right)
\end{eqnarray} 
Using this expression in the expression for $\lambda N$ above, we get:
\begin{eqnarray} 
\lambda N &=& \frac{1}{3\pi^2} \left(\frac{2m}{\hbar^2}\right)^{3/2} \int d\vec{r}\,
\left\{\left(\lambda - V\right)^{5/2}\right\} \Theta\left(\lambda - V\right) \nonumber \\
     &+& \frac{1}{3\pi^2} \left(\frac{2m}{\hbar^2}\right)^{3/2} \int d\vec{r} \, 
\left\{V\left(\lambda - V\right)^{3/2}\right\} \Theta\left(\lambda - V\right) \nonumber \\
\end{eqnarray}  
up to leading order. This expression, when combined with that for $E_{\hbar^{0}}^{CN}$, yields:
\begin{eqnarray} 
E_{TF} &=& \frac{1}{3\pi^2}\left(\frac{2m}{\hbar^2}\right)^{3/2}\int d\vec{r}\, \left\{\frac{3}{5} 
\left(\lambda - V\right)^{5/2}\right\}\Theta\left(\lambda - V\right) \nonumber \\
       &+& \frac{1}{3\pi^2}\left(\frac{2m}{\hbar^2}\right)^{3/2}\int d\vec{r}\,
\left\{ V\left(\lambda - V\right)^{3/2}\right\}\Theta
\left(\lambda - V\right) \nonumber \\ 
\end{eqnarray} 
On the right hand side of $E_{TF}$, the first term corresponds to the kinetic energy and the 
second to the potential energy within the Thomas Fermi approximation. Proceeding along the same 
lines, we get the usual expressions (see, for example, \cite{brack97}) for second order corrections to kinetic 
and potential energies (excluding the spin-orbit contributions):
\begin{eqnarray} 
E_{\hbar^2}^{\rm KE} &=& \frac{1}{24\pi^2}\left(\frac{2m}{\hbar^2}\right)^{1/2}
\int d\vec{r}\left\{ \vphantom{\frac{x^5}{x^5}} \left(\lambda - V\right)^{1/2}\nabla^{2}V \right. \nonumber \\
       && \left. - \frac{3}{4}\frac{\left(\nabla V\right)^2}{
\left(\lambda - V\right)^{1/2}} \right\} \Theta\left(\lambda - V\right) \\
E_{\hbar^2}^{\rm PE} &=& \frac{-1}{24\pi^2}\left(\frac{2m}{\hbar^2}\right)^{1/2}
\int d\vec{r}\left\{ \frac{V \nabla^{2}V } {\left(\lambda - V\right)^{1/2}} \right. \nonumber \\
       && \left. + \frac{1}{4}\frac{V \left(\nabla V\right)^2}{
\left(\lambda - V\right)^{3/2}} \right\} \Theta\left(\lambda - V\right)
\end{eqnarray} 

\vskip\baselineskip 
\appendix{{\bf Appendix 2: Explicit Formulas for Derivatives of Modified Woods - Saxon Form Factor}}

\vskip\baselineskip 
Here we list the explicit formulas for derivatives of Woods - Saxon form factor. Let 
\begin{eqnarray} 
g(r) = \Bigl[g_{0}\left(\vec{r}\right)\Bigr]^{\gamma}
\end{eqnarray} 
with $\gamma > 0$ and  
\begin{eqnarray} 
g_{0}\left(\vec{r}\right) = \frac{1}{1 + \exp\left(l\left(\vec{r}\right)/a\right)}
\end{eqnarray} 
with $l\left(\vec{r}\right)$ is suitably defined distance function (see, for example, 
\cite{bhagwat10}) and $a$ is the diffusivity parameter. We shall henceforth suppress all the 
arguments of the functions for the sake of brevity. Define:
\begin{eqnarray} 
\xi &=& g\left(g_{0} - 1\right) \\
\xi_0 &=& g_0\left(g_{0} - 1\right) \\
\zeta &=& \left(\gamma + 1\right)g_{0} - \gamma
\end{eqnarray} 
Evaluation of the different contributions to the semi - classical energy requires 
gradient of $g$ as well as derivatives of $\left(\nabla g\right)^2$. These are listed
below first. We have:
\begin{eqnarray} 
\nabla g = \frac{\gamma}{a}\, \xi \nabla l
\end{eqnarray} 
giving us,

\begin{widetext}
\begin{eqnarray} 
\left(\nabla g\right)^{2} &=& \frac{\gamma^2}{a^2}\, \xi^2 \left(\nabla l\right)^{2} \\
\nabla\left(\nabla g\right)^{2} &=& \frac{\gamma^2}{a^2}\,\xi^2 \left[\nabla\left(\nabla l\right)^2 
      + \frac{2}{a}\,\zeta\left(\nabla l\right)^2 \nabla l\right] \\
\nabla^2\left(\nabla g\right)^2 &=& \frac{2\gamma^2}{a^3}\,\xi^2\,\zeta\left[\nabla l\cdot\nabla\left(\nabla l\right)^2
 + \frac{2}{a}\,\zeta\left(\nabla l\right)^{4}\right] 
+ \frac{\gamma^2}{a^2}\,\xi^2\left[\nabla^2\left(\nabla l\right)^2 + \frac{2}{a}\,\zeta\left\{\nabla^2 l 
\left(\nabla l\right)^2 + 
 \nabla l \cdot \nabla\left(\nabla l\right)^2\right\}  \right. \nonumber \\ 
&& \left. + 
\frac{2}{a^2}\left(\gamma + 1\right)\xi_0\left(\nabla l\right)^{4}\right]
\end{eqnarray} 
\end{widetext}

In addition to these, the expressions for different terms appearing in the semi - classical energy
requires Laplacian of $g$ as well as derivatives of $\nabla^2 g$. These can be expressed as:

\begin{widetext}
\begin{eqnarray} 
\nabla^2 g &=& \frac{\gamma}{a}\,\xi\left[\nabla^2 l + \frac{1}{a}\,\zeta\left(\nabla l\right)^2\right] \\
\nabla\left(\nabla^2 g\right) &=& \frac{\gamma}{a^2}\,\xi\,\zeta\left[\nabla^2 l 
+ \frac{1}{a}\,\zeta\left(\nabla l\right)^2\right]\nabla l 
+ \frac{\gamma}{a}\,\xi\left[\vphantom{\frac{1}{1}}\nabla\left(\nabla^2 l\right)
+ \frac{1}{a^2}\left(\gamma + 1\right)\xi_0
\left(\nabla l\right)^2 \nabla l + \frac{1}{a}\,\zeta\nabla\left(\nabla l\right)^2\right] \\
\nabla^2\left(\nabla^2 g\right) &=& \frac{\gamma}{a^2}\,\xi\,\zeta\left[\nabla^2 l 
+ \frac{1}{a}\,\zeta\left(\nabla l\right)^2\right]\nabla^2 l 
+ \frac{2\gamma}{a^2}\,\xi\,\zeta\left[\nabla l \cdot\nabla\left(\nabla^2 l\right) 
+ \frac{1}{a}\,\zeta\nabla l \cdot\nabla\left(\nabla l\right)^2 
+ \frac{1}{a^2}\left(\gamma + 1\right)\xi_0\left(\nabla l\right)^4\right] \nonumber \\
&+& \frac{\gamma}{a^3}\,\xi\left(\nabla l\right)^2 
\left\{\vphantom{\frac{1}{1}} \zeta^2 + \left(\gamma + 1\right)\xi_0\right\}
\left[\nabla^2 l + \frac{1}{a}\,\zeta\left(\nabla l\right)^2\right]
+ \frac{\gamma}{a}\,\xi\left[\nabla^2\left(\nabla^2 l\right) + \frac{1}{a^3}\left(\gamma + 1\right)
\left(2g_0 - 1\right)\xi_0\left(\nabla l\right)^4 \right. \nonumber \\ 
&& \left. + \frac{1}{a^2}\left(\gamma + 1\right)\xi_0\left(\left(\nabla l\right)^2 \nabla^2 l 
+ 2\nabla l \cdot \nabla\left(\nabla l\right)^2\right)  + \frac{1}{a}\,\zeta\nabla^2\left(\nabla l\right)^2\right]
\end{eqnarray} 
\end{widetext}

The derivatives of distance function are not presented here, since the details would depend on 
the particular choice of distance function. In the present context, we are assuming reflectionally
symmetric shapes. Therefore, it is most convenient to work in spherical polar system, as 
explained in \cite{bhagwat10}. In this case, spherical harmonics appearing in the distance 
function can be suitably combined to write $R_s$ (see Eq. (6)) in terms of 
multiple angle formulas of the trigonometric functions, which makes explicit evaluation of the
derivatives relatively easier. 

\begin{acknowledgments}
A.B. is thankful to Departament de F\'isica Qu\`antica i Astrof\'isica and Institut 
de Ci\`encies del Cosmos, Facultat de F\'isica, Universitat de Barcelona for their kind hospitality.
M.C. and X.V. were partially supported by Grant FIS2017-87534-P from MINECO and FEDER, and by
Grant CEX2019-000918-M from the State Agency for Research of the Spanish Ministry of Science and Innovation
through the Unit of Excellence Mar\'{\i}a de Maeztu 2020-2023 award to ICCUB.

\end{acknowledgments}

\end{document}